\pdfoutput=1
\documentclass[table, xcdraw, sigconf]{acmart}
\acmConference[ESEC/FSE 2017]{11th Joint Meeting of the European Software Engineering Conference and the ACM SIGSOFT Symposium on the Foundations of Software Engineering}{4--8 September, 2017}{Paderborn, Germany}
\usepackage{color}
\usepackage{balance}
\usepackage{xcolor}
\usepackage{multirow}
\usepackage{booktabs} 
\usepackage[export]{adjustbox}
\usepackage{url}
\usepackage{hhline}
\usepackage{amssymb}
\usepackage{amsfonts}
\usepackage{amsmath}
\usepackage{hyperref}
\usepackage{algorithm}
\usepackage{algorithmic}
\usepackage{pbox}

\newcommand{\overbar}[1]{\mkern 1.5mu\overline{\mkern-1.5mu#1\mkern-1.5mu}\mkern 1.5mu}

\definecolor{Gray}{gray}{0.95}
\definecolor{LightGray}{gray}{0.975}

\newcommand{\bi}{\begin{itemize}[leftmargin=0.4cm]}
	\newcommand{\ei}{\end{itemize}}
\newcommand{\be}{\begin{enumerate}}
	\newcommand{\ee}{\end{enumerate}}

\newcommand{\fig}[1]{Figure~\ref{fig:#1}}

\colorlet{shadecolor}{blue!20}

\definecolor{Gray}{rgb}{0.88,1,1}
\definecolor{Gray}{gray}{0.85}
\makeatletter
\let\th@plain\relax
\makeatother
\usepackage[framed]{ntheorem}
\usepackage{framed}
\usepackage{tikz}
\usetikzlibrary{shadows}

\theoremclass{insight}
\theoremstyle{break}
\tikzstyle{thmbox} = [rectangle, rounded corners, draw=black,
 fill=Gray,  drop shadow={fill=black, opacity=1}]

\newshadedtheorem{insight}{Insight}

\setcopyright{rightsretained}

\acmDOI{XX.YY/ZZ}

\acmISBN{ZZ-YY-24-ZZ/QQ/A}

\acmConference[FSE'17]{Germany}{Sept 2017}{} 
\acmYear{2017}
\copyrightyear{2017}

\acmPrice{15.00}

\begin{document}
\title{Replicating and Scaling up Qualitative Analysis using Crowdsourcing: A Github-based Case Study}

\author{Di Chen, Kathryn T. Stolee, Tim Menzies}
\affiliation{%
  \institution{North Carolina State University, USA}
  \city{Raleigh} 
  \state{NC} 
  \postcode{27606}
}
\email{dchen20@ncsu.edu,  ktstolee@ncsu.edu,  tim@menzies.us}




\begin{abstract}
	Due to the difficulties in replicating and scaling up qualitative studies, 
	such studies are rarely verified.
	Accordingly, in this paper, we leverage the advantages of crowdsourcing 
	(low costs, fast speed, scalable workforce) 
	to replicate and scale-up one state-of-the-art qualitative study.
	That qualitative study explored   20 GitHub pull requests to learn
	factors that  influence the fate of pull requests with respect to approval and merging.
	
	As a secondary study, using crowdsourcing 
	at a cost of \$200, we studied 250 pull requests from 142 GitHub projects.
	The prior qualitative findings are mapped into questions for crowds workers. Their answers were converted into binary features to build a predictor which predicts whether code would be merged with median $F_1$ scores of 68\%. For the same large group of pull requests, the median $F_1$ scores could achieve 90\% by a predictor built with additional features defined by  prior quantitative results.

Based on this case study, we conclude that
  there is much benefit in combining  different kinds of research methods.
  While qualitative insights are very useful for finding novel insights, 
they can be  hard to scale or replicate. That said, they can guide and define the goals
of scalable secondary studies that use (e.g.) crowdsourcing+data mining.
On the other hand,  while data mining  methods   are reproducible
and scalable to large data sets, their results may be spectacularly wrong
since they lack contextual information.
That said, 
they can be used to test the stability and external validity,  of the insights
gained from a qualitative analysis. 
\end{abstract}
 
%
%


\maketitle

\section{Introduction}

Our ability to generate models from software engineering (SE) data has out-paced our abilities
to reflect on those models.
Studies can use 
thousands of projects, 
millions of lines of code, 
tens of thousands of programmers
~\cite{ray2014large}.
But
when insights from human experts
are overlooked, the conclusions from the
automatically generated models can be both wrong and
misleading~\cite{o2016weapons}.  
After observing 
case studies where data mining in SE
led to spectacularly wrong results, Basili and Shull~\cite{shull02} 
recommend qualitative analysis to collect and use insights from subject
matter experts who understand software engineering.

Shull~\cite{shull13}  also warns that  traditional methods of finding local beliefs
 (based on an anthropological-style 
 analysis) does not scale  to
 hundreds of projects.
 Shull notes
 that those manual methods have trouble keeping up with the  pace of technological change.
Such studies can take years to complete -- in which time, the underlying
technology may have completely changed.

To solve this problem, we propose   {\em scalable secondary studies}. Such 
studies are conducted
after collecting
  qualitative insights  from an in-depth analysis of a small sample.
Next, other method(s) are employed to ensure conclusion stability
and external validity in a larger sample.
Such scalable secondary studies can be implemented
using a variety of methods; in this work, we explore crowdsourcing+data mining.

To demonstrate this approach, we extend
a study from   FSE'14  by   Tsay,   Dabbish,     Herbsleb (hereafter, TDH)~\cite{tsay2014let}. TDH
explored how
Github-based teams debate what new
code gets merged.
To do this, they used a 
labor-intensive qualitative interview-process of 47 users of GitHub, 
as well as in-depth case studies of 20 pull-requests.

This papers extends  that  primary qualitative study of pull requests with a scalable secondary
study using crowdsourcing+data mining.  Using the terms identified by TDH,   crowdsourcing
  extracted data from 250 pull requests (i.e. an order of magnitude more that TDH).
Data mining was applied to that new data resulting in
  accurate predictors for generating what issues will get merged.
Further, when those predictors were compared to another predictor
built using more quantitative methods (traditional data mining, no crowd-sourcing,
no use initial qualitative insights), this second predictor out-performed
the TDH features (the $F_1$ score
grew from 68\% to 90\%).  

This is not to say that  data mining methods are ``better''
than qualitative methods. In fact, our key point is:
\begin{quote}
{\em
This  secondary studies could not
have have happened  \underline{without} the initial
qualitative study.}
\end{quote}
That, in this work,
the qualitative inspired and guided the
subsequent work.
Specifically,  
the TDH qualitative study defined (a)~baseline results and (b)~ a challenge task
addressed  in a subsequent secondary study that used crowdsourcing+data mining.
In this case, that secondary study is able to build predictors using larger amount of pull requests analyzed.
But it would be extremely premature 
to use this one result to make some general judgement about the relative merits
of different approaches. As James Herbsleb said in his FSE'16 keynote
address, ``these methods exist and we need to learn to best use them all''.
Accordingly, we recommend   several years of work where researchers explore
  combinations of methods.  This paper will be a success if it
  encourages more  studies on  combinations of primary
  qualitative studies and subsequent secondary studies.

The contributions of this work are:
\begin{itemize}
    \item A cost-effective, independent replication of a primary study of pull request acceptance factors using a scaled sample of artifacts (RQ1). 
    \item Analysis of the external validity of the original study, demonstrating stability in some of the results. This has implications for which questions warrant further analysis (RQ2). 
    \item A literature review on factors that impact pull request acceptance and identification of features that reliably predict pull request acceptance (RQ3). 
\end{itemize}

The rest of this paper details our \underline{M}ethodology f\underline{O}r \underline{S}calable 
\underline{S}econdary 
\underline{S}tudies (MOSSS).
    The next  section introduces 
    empirical methods in software engineering;
    compares qualitative and quantitative methods in software engineering;
    describes GitHub and pull requests;
    and offers an overview on crowdsourcing.
    Next, in the \textit{Methods} section, we describe 
    the details of how we apply MOSSS on TDH.
    Our \textit{Results} section presents our findings. 
    This is followed by \textit{Threats to Validity} and \textit{Conclusions}.

        
    
            	
        	
    

    To assist other researchers, a reproduction package with all
    our scripts and data is available in GitHub\footnote{\url{https://github.com/dichen001/FSE_17}} and in archival
    form, tagged with a DOI\footnote{\url{https://doi.org/10.5281/zenodo.322925}} (to simplify all future citations to this material).


\section{Background and Related Work}

    \subsection{Empirical SE Methods}
    There are many ways to categorize empirical studies in SE. 
    Sjoberg, Dyba et. al.~\cite{sjoberg2007future} summarize them into two general groups, i.e. primary research and secondary research.
    The most common primary research 
    usually involves the collection and analysis of original
    data, utilizing methods such as experimentation,
    surveys, case studies, and action research. 
    While our paper falls into the secondary research that 
    uses data from previously published
    studies for the purpose of research synthesis,
    which involves summarizing, integrating and combining
    the findings of different studies on a research question~\cite{cooper2009handbook}.
    According to Cohen~\cite{cohen1989developing}, secondary studies can identify
    crucial areas and questions that have not been addressed
    adequately with previous empirical research. 
    The core observation it is built on is that no matter how well designed
    and executed, findings from single empirical
    studies are limited in the extent to which they may be
    generalized. 
    
    For the methods in empirical SE studies, we have participant observation, interviewing and coding for data collection, constant comparison and cross-case analysis for theory generation and replication for theory confirmation~\cite{seaman1999qualitative}. 
    Our data collection applies the coding method, which is commonly used to extract values for
    quantitative variables from qualitative data in order to perform some
    type of quantitative or statistical analysis. 
    Our data analysis falls into the category of replication for theory confirmation. Using replication, we could check the external validity, which focuses on whether claims for the generality of TDH results are
    justified, and reliability, which focuses on whether our study yields the same results if we replicate it~\cite{easterbrook2008selecting}.

    \subsection{Qualitative, Quantitative, and Mixed-Methods in SE}
	{\em Qualitative methods} in SE are typically used for exploratory
        purpose to generate new theorems or improve existing
        ones~\cite{sale2002revisiting, lazaro2006approach}. Due to
        the involvement of humans when qualitative method are applied,
        the sample size is often restricted to a very small size.
        Additionally, the results of a 
        qualitative analysis can be  difficult to replicate due to variations in settings or experimenter bias, 
        and therefore engaging in such tasks are risky to researchers~\cite{Shull:2008:RRE:1361580.1361587, guba1994competing}.

	In contrast, {\em quantitative studies} are used for explanatory
        or descriptive purpose to measure and analyze causal
        relationships between variables~\cite{sale2002revisiting, lazaro2006approach}. Quantitative methods work with
        numerical data collected from a representative sample, which
        could be very large compared to qualitative methods.    
        
    A drawback with quantitative
    methods is that they may operate
    without any contextual knowledge
    of the data they are processing. A common study in the mining
    software repositories community is to apply data miners on information
    taken from some
    repository without first interviewing humans familiar with that project
    or that data. 
    Hence, such quantitative methods
    may efficiently reach a conclusion
    over a very large data set,
    even though those conclusions may not address any current concerns of any
    living human.
    
    For these reasons, various researchers explore  
	{\em mixed methods}  to exploit the strengths of all the above approaches. 
	For example, Zimmermann and his colleagues at Microsoft
	conduct very focused limited-scope interviews with a small number of developers.
	This primary qualitative analysis is used to refine a set of hypotheses and questions for a secondary
	study comprising a questionnaire distributed to a very
	wide audience~\cite{begel2014analyze}.
		    The results of that questionnaire are then summarised using a ``card sort'' (which, we note, is a method commonly used by researchers in interpreting free text responses in survey data ~\cite{stoleeFSE2015, bacchelli2013expectations, guzzi2013communication, siegmund2015views}).
	This is a very labor-intensive and somewhat subjective process whereby researchers work through all the questionnaire textual answers,
	organizing the  topics into categories. To the best of our knowledge, even though card sorting is typically done by multiple researchers who reach a consensus, 
	the results of an initial card sort are typically not verified by
	a second card sort with independent researchers. We speculate that card sorts are such a resource-intensive
	undertaking that doing it twice is just impractical.
	
	The proposal in this paper is that mixed methods
	can be improved.
	Primary studies should remain qualitative since their careful
	and detailed analysis of human factors within a software project
	is insightful. Also, they can lead to novel insights they can be overlooked
	by automatic data mining methods. However, once the primary qualitative
	study is completed, the stability and external validity of those
	conclusions should be checked by a scalable secondary method, ideally by
	an independent research team~\cite{Shull:2008:RRE:1361580.1361587}.
	One example of such a scalable secondary method is the combination
	of data mining+crowd sourcing explored in this paper.

\subsection{GitHub and Pull Requests}

    \begin{figure}[!t]
			\centering
			\includegraphics[width=3.1in]{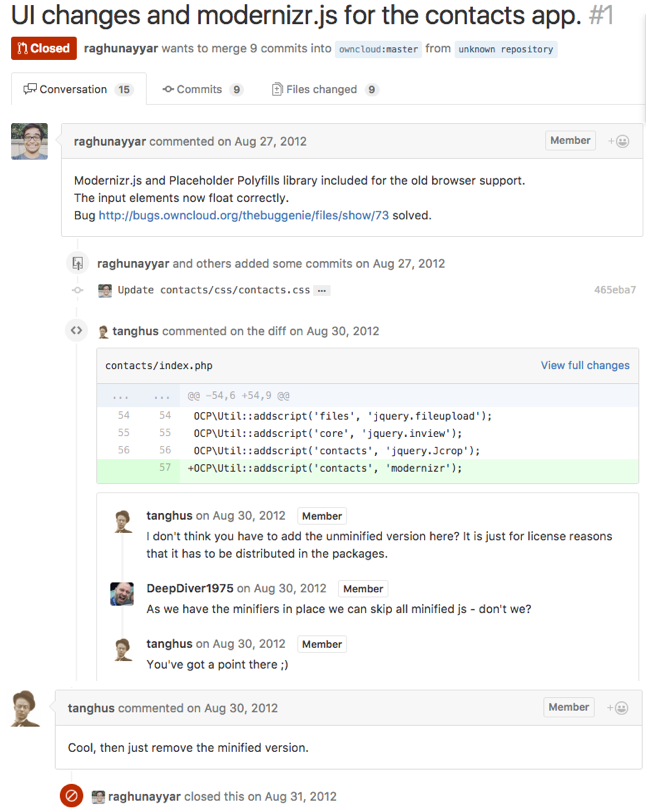}
			\caption{An example GitHub pull request}
			\label{fig:pr}
		\end{figure}
    
    With over 14 million users and 35 million repositories as of April 2016, GitHub has become the largest and most influential open source projects hosting sites.
    Numerous qualitative ~\cite{ dabbish2012social, tsay2014let, gousios2015work, gousios2016work, marlow2013impression, begel2013social,mcdonald2013performance, pham2013creating}, and  
    quantitative ~\cite{tsay2014influence, gousios2013exploration, tsay2012social, yu2015wait, zhang2014investigating, rahman2014insight, takhteyev2010investigating, thung2013network, ray2014large} 
    and mixed methods studies ~\cite{kalliamvakou2014promises, blincoe2016understanding} 
    have been published about GitHub. 

    Pull requests need to be created when 
    contributors want their changes to be merged to the main repository. 
    After core members receive pull requests, 
    they inspect the changes and 
    decide whether to accept or reject them. 
    This process usually involves code inspection, 
    discussion and inline comments between contributors and repository owners. 
    Core members who have the ability to close the pull requests 
    by either accepting the code and merging the contribution with the master branch, 
    or rejecting the pull requests. 
    Core members could also ignore the pull requests and leave them in the open state.  
    Figure ~\ref{fig:pr} shows an example of pull requests with discussion, inline code comments and final result.

\subsection{Crowdsourcing}

        One of first uses of this term  comes from Jeff Howe in 2006\footnote{\url{http://www.crowdsourcing.com/cs/2006/06/crowdsourcing_o.html}} who said:
        \begin{quote}
        "crowdsourcing represents the act of a company or institution taking a function once performed by employees and outsourcing it to an undefined (and generally large) network of people in the form of an open call."
        \end{quote}        
        \noindent

Crowdsourcing is being used for many  SE tasks including
        program synthesis~\cite{Cochran:2015:PBP:2676726.2676973},  
        program verification~\cite{Schiller:2012:RBW:2398857.2384624}, and  
        testing~\cite{6569745, Nebeling:2013:CGT:2494603.2480303} using tools such as
 Amazon Mechanical Turk (MTurk)\footnote{MTurk focuses on micro-tasks, e.g., labeling an image. Micro-tasks are grouped into one Human Intelligence Task (called HIT). When HITs are defined, they can include the HIT payment, the time constraint for answering, the expiration time for a job to be available on MTurk, and some qualification test. For more, see \url{https://www.mturk.com/mturk/welcome}.} 
        TopCoder\footnote{TopCoder is a platform designed to use crowdsourcing  for large software engineering tasks, such as website design or implementation. Designers, employed by TopCoder, break down large SE projects into small tasks for crowd workers. For more, see \url{https://www.topcoder.com}.}, 
        CrowdFlower\footnote{CrowdFlower is somewhat similar to MTurk but adds some quality control micro-tasks. At CrowdFlower, 20\% of the micro-tasks assigned to responders are ``golden''; i.e. the answers are already known. For more, see \url{http://www.crowdflower.com}.},   
        and ClickWorker\footnote{ClickWorker is similar to MTurk and CrowdFlower in that it's a micro-task crowdsourcing platform, but the focus is on surveys, proofreading, web research. Skills are measured by ClickWorker {\em before} workers have access to actual tasks. For more, see \url{https://www.clickworker.com}.}. For an extensive review of crowdsourcing in software engineering, see~\cite{mao2015survey},
 
    \begin{figure}[!t]
		\center
		\includegraphics[width=2.1in]{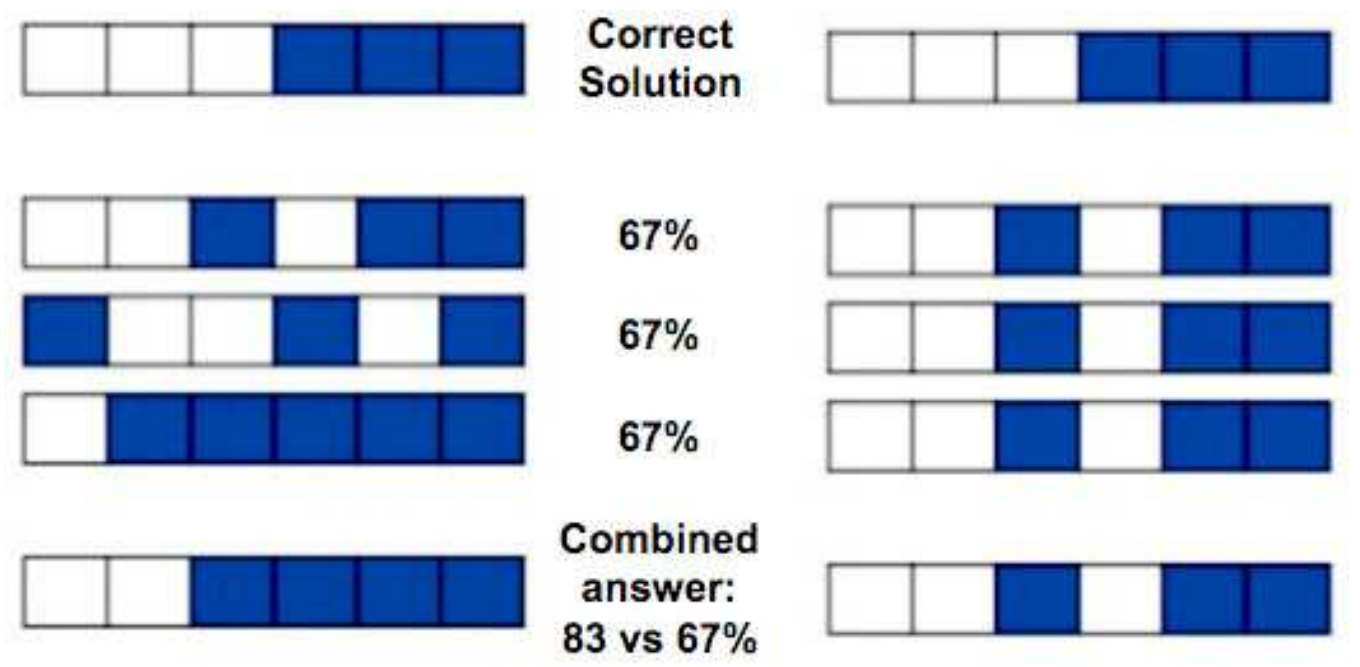}
		\caption{Combing crowd opinions (at left) can work better 
		than using individual opinions (at right). Here, the fifth line
		averages the opinion of three individuals. Credit:~\cite{minku15}.}
		\label{fig:minku}
	\end{figure}
	
  \begin{figure*}[ht]
        \centering
        {\includegraphics[width=4.6in]{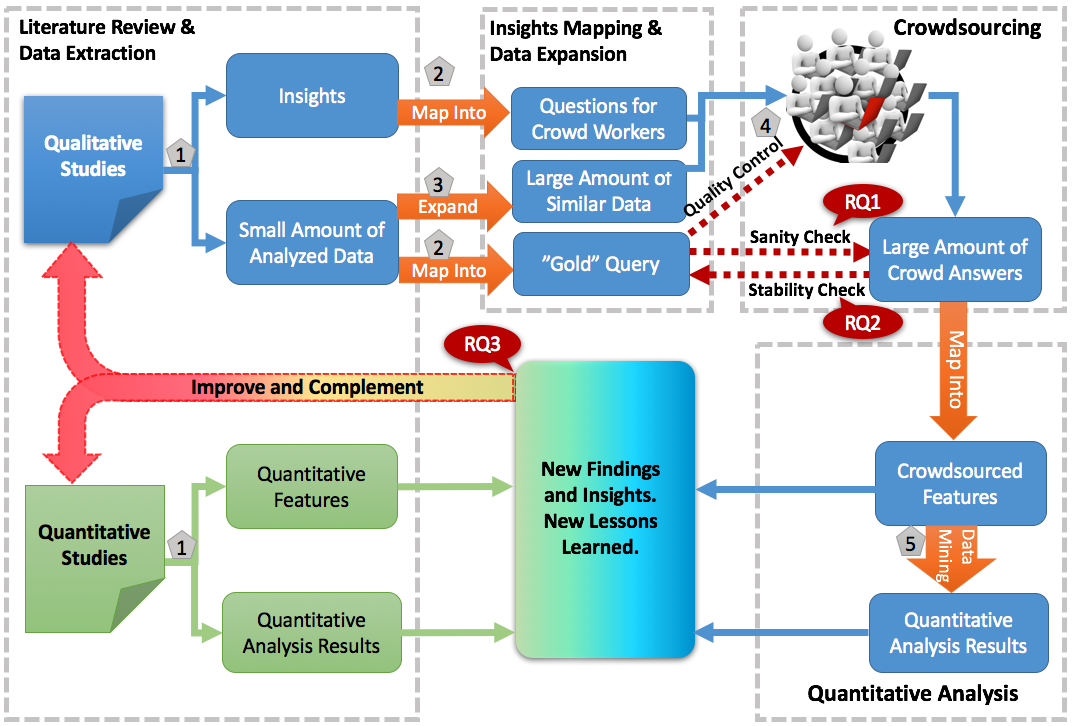}}
	    \caption{Methodology of Scalable Secondary Studies (MOSSS).
	    Note the red bubbles
	    denoting the location
	    of our three research questions
	    \textcolor{red}{{\bf RQ1, RQ2, RQ3}}
	    within this framework.}
		\label{fig:MOSSS}
    \end{figure*}

There are many reasons to use crowdsourcing. Firstly, it is very cheap.
As described later in this paper, this entire study requires just \$200 of crowd time.
Secondly, crowdsourcing can solve  some very hard problems. 
For example,  Minku~\cite{minku15} argues from sampling theory that 
    the union of $N$ slightly different views can be larger than any individual view (see \fig{minku}). 
    For example, in   software engineering studies, it was found that 
      combining crowd answers for writing regular expressions for URLs or email addresses 
    yield better accuracy programs written by a single individual~\cite{Cochran:2015:PBP:2775051.2676973}. 
    
    Surprisingly,  the crowd has an unexpectedly coherent understanding of its distributed knowledge. For
    example, Heikinheimo and Ukkonen's {\em centroid median} theorem shows that if a crowd checks numerous examples for outliers, then   items   marked least  are equal to the  mean of a univariate normal distribution~\cite{Heikinheimo2013}. 
    Note that centroid medians work for single numbers in a range or sets of large complex feature vectors.
    That is, crowds can be used as a human-based data miner to implement, 
    for example, a crowd-sourced K-means clusterer (as done by  Heikinheimo and Ukkonen). 

    Significantly,  the crowd can solve problems that have defeated conventional computer science methods. Mason~\cite{Mason2013} comments that the larger the crowd, 
    the more likely that an individual will offer a (possibly partial) solution to even intractable NP-hard problem. If any  partial solution is offered, due to the size of the crowd, it becomes easier to confirm
    or refute the solution~\cite{amir2013verification}.

Two important issues with crowdsourcing are        {\em quality control} and {\em cost control}.
 Crowd-based workers may return noisy or
        incorrect results. One way to address this issue is to use, say, 20\%  ``gold''
        questions-- those for which the answer is known。 Workers that perform poorly on the golden set are eliminated, which is also one of the strategies we take in this paper. Alternatively, tasks can be assigned to multiple workers and their results aggregated (e.g., in the TURKIT system~\cite{5295247}, one task
        is performed iteratively, and each worker is asked to improve on the answer of the former, or in AutoMan~\cite{Barowy:2012:API:2398857.2384663}, each task is performed by crowd members until statistical consensus is reached). Other quality control techniques include redundant question formats~\cite{stolee2015exploring}, notifications to workers about the value of their work~\cite{krosnick1991response}, answer conflict detectors~\cite{shiel2013conflict}, and random click detection~\cite{Kim:2012:FOR:2442576.2442591}.
        
        As to cost control, commercial crowdsourcing platforms are not free. Economic incentives for crowd workers strongly effect  crowd response quality~\cite{mason2010financial, goel2014mechanism, mao2013volunteering, kittur2008crowdsourcing, mason2011use,Mao:2013:PCS:2486788.2486963, yin2014monetary, wang2013quality}. To keep the quality high, payments need to be high enough to entice participation~\cite{vinayak16}.

  \section{Research Questions}
    In this work, we evaluate the following research questions: 
    
      \noindent {\bf  RQ1:} {\em Can we use crowdsourcing to replicate TDH's work, but at faster speeds, with cheaper costs and larger sample sizes?}
       According to Shull's notes
       on how to define   replication studies in SE\cite{Shull:2008:RRE:1361580.1361587}, in order to show that a given result is robust, the ideal case is for a completely independent set of researchers to replicate a published study using their own experimental design. In this paper, our analysis, which is designed and run by a completely independent set of researchers compared to the primary study, scales well and cost-effectively  validates and extends the prior TDH study.\\ 
       
       
       
	  \noindent {\bf  RQ2:} {\em Does crowdsourcing lead to stable conclusions? } 
	  This question is important since, given the nature of crowd-based reasoning, it is possible that crowd workers will have different opinions from TDH. We collected the 210 pull requests using similar sampling criteria to the primary study and tested if the crowd reaches the same or different conclusions using the new data set.  A conclusion was declared stable if the same conclusions were found in each independent sample. As shown, the analyses in this paper are stable across these two samples. \\
	  
	  \noindent {\bf  RQ3:} {\em Can the pull request features identified in the primary study accurately predict pull request acceptance? } 
	  Given the larger data set collected and evaluated in this work, there is now an opportunity to evaluate the performance of prediction models based on (1) the features identified as important in the primary study, and (2) features identified as important in previous data mining-only studies.


    \begin{table*}[t!]
\centering
\caption{A sample of related qualitative and quantitative work.
Here, by ``quantitative'', we mean using data mining with little
to no interaction with project personnel.}
  \label{tab:lit_rev}
\begin{adjustbox}{max width=6.5in}
\begin{tabular}{|l|l|l|l|l|}
\hline
\rowcolor[HTML]{C0C0C0} 
\textbf{Year}                     & \textbf{Source} & \textbf{Method}              & \textbf{Data}                                                          & \textbf{Title}                                                                                  \\ \hline
2012~\cite{dabbish2012social}        & CSCW            & Qualitative                  & Interview 24 GitHub Users.  Pull requests case study 10.               & Social coding in GitHub: transparency and collaboration in an open software repository          \\ \hline
2013~\cite{marlow2013impression}     & CSCW            & Qualitative                  & Interview 18 GitHub users.  Pull requests case study 10.               & Impression Formation in Online Peer Production: Activity Traces and Personal Profiles in GitHub \\ \hline
2014~\cite{tsay2014let}              & FSE             & Qualitative                  & Interview 47 GitHub users.  Pull requests case study 20.               & Let's Talk About It: Evaluating Contributions through Discussion in GitHub                      \\ \hline
2015~\cite{gousios2015work}          & ICSE            & Qualitative                  & Online survey 749 integrators.                                         & Work Practices and Challenges in Pull-Based Development: The Integrator’s Perspective           \\ \hline
2016~\cite{gousios2016work}          & ICSE            & Qualitative                  & Online survey 645 contributors.                                        & Work Practices and Challenges in Pull-Based Development: The Contributor’s Perspective          \\ \hline
                                  &                 &                              &                                                                        &                                                                                                 \\ \hline
2014~\cite{gousios2014exploratory}   & ICSE            & Quantitative                 & GHTorrent, 166,884 pull requests                                       & An Exploratory Study of the Pull-Based Software Development Model                               \\ \hline
2014~\cite{tsay2014influence}        & ICSE            & Quantitative                 & GitHub API, GitHub Archive. 659,501 pull requests                      & Influence of Social and Technical Factors for Evaluating Contribution in GitHub.                \\ \hline
2014~\cite{yu2014reviewer}           & ICSME           & Quantitative                 & GHTorrent, 1,000 pull requests.                                        & Reviewer Recommender of Pull-Requests in GitHub                                                 \\ \hline
2014~\cite{vasilescu2014continuous}  & ICSME           & Quantitative                 & GHTorrent                                                              & Continuous Integration in a SocialCoding World Empirical Evidence from GitHub                   \\ \hline
2014~\cite{yu2014should}             & APSEC           & Quantitative                 & GHTorrent, 1,000 pull requests.                                        & Who Should Review This Pull-Request: Reviewer Recommendation to Expedite Crowd Collaboration    \\ \hline
2014~\cite{zhang2014investigating}   & CrowdSoft       & Quantitative                 & GHTorrent, GitHubArchive.                                              & Investigating Social Media in GitHub’s Pull-Requests: A Case Study on Ruby on Rails             \\ \hline
2014~\cite{gousios2014dataset}       & MSR             & Quantitative                 & GHTorrent                                                              & A Dataset for Pull-Based Development Research                                                   \\ \hline
2014~\cite{rahman2014insight}        & MSR             & Quantitative                 & GHTorrent,  78,955 pull requests.                                      & An Insight into the Pull Requests of GitHub                                                     \\ \hline
2014~\cite{pletea2014security}       & MSR             & Quantitative                 & GHTorrent,  54,892 pull requests.                                      & Security and emotion sentiment analysis of security discussions on GitHub                       \\ \hline
2014~\cite{brunet2014developers}     & MSR             & Quantitative                 & GHTorrent                                                              & Do developers discuss design                                                                    \\ \hline
2014~\cite{padhye2014study}          & MSR             & Quantitative                 & GHTorrent, 75,526 pull requests.                                       & A study of external community contribution to opensource projects on GitHub                     \\ \hline
2015~\cite{van2015automatically}     & MSR             & Quantitative                 & GHTorrent                                                              & Automatically Prioritizing Pull Requests                                                        \\ \hline
2015~\cite{yu2015wait}               & MSR             & Quantitative                 & GHTorrent, 103,284 pull requests.                                      & Wait For It: Determinants of Pull Request Evaluation Latency on GitHub                          \\ \hline
                                  &                 &                              &                                                                        &                                                                                                 \\ \hline
2014~\cite{kalliamvakou2014promises} & MSR             & \pbox{20cm}{ Quantitative \& \\ Qualitative}  & \pbox{20cm}{ Quant. : GHTorrent \\ Qual. : 240 Survey, Manual analysis 434 project.} & The promises and perils of mining GitHub                                                        \\ \hline
\end{tabular}
\end{adjustbox}

\end{table*}

\section{Methods}

    To leverage the advantages of crowdsourcing, we propose our methodology of scalable secondary studies (MOSSS) for quickly replicating and scaling the time-consuming qualitative works in SE. Our methodology will be introduced and applied on the primary study from TDH on GitHub pull requests. Basically, our methodology shown in Figure~\ref{fig:MOSSS} is consist of the following steps:
    
    \noindent {\bf Step 1:} {\em Literature review on one specific domain, e.g. GitHub pull requests studies, and extract data, insights, features and results from the existing work.}
    
    \noindent {\bf Step 2:} {\em Map insights from qualitative works into questions that could be easily answered by crowd workers and quantitative features should also be easily extracted from these questions. Similarly, map existing data into questions with known answers, which are `gold' queries.}
    
    \noindent {\bf Step 3:} {\em Expand the data used in the primary studies with similar selection rules and launch some initial data on crowdsourcing for cost control.}
    
    \noindent {\bf Step 4:} {\em Apply the mapped questions with crowdsourcing on the expanded data, while using "gold" queries for quality control in crowdsourcing.}
    
    
    
    
    \noindent {\bf Step 5:} {\em Extract features defined in Step 2 from the large amount of answers returned by crowd workers, then apply quantitative analysis on these crowdsourced features and compare those with the quantitative features in Step 1 so as to discover new findings.}

    
    \subsection{Literature Review and Data Extraction}

    We first set TDH as our primary study and then find all studies related to GitHub pull requests in the literature.    
    We then searched keywords `pull', `request' and `GitHub' on Google Scholar since 2008 and a dataset from ~\cite{mathew17}, which contains 16 software engineering conferences, 1992 to 2016, which includes \textit{ICSE, ICSM, WCRE, CSMR, MSR, GPCE, FASE, ICPC, FSE, SCAM, ASE, SANER, SSBSE, RE', ISSTA, ICST}. 
    After manually reviewing the search results, we filtered out the work unrelated to GitHub pull requests. 
    Table~\ref{tab:lit_rev} lists the remaining research papers that have studied pull requests in GitHub using either qualitative or quantitative methods. 
    Here, we distinguish qualitative and quantitative methods by whether or not there is human involvement during data collection process. Qualitative studies have human involvement and include interviews, controlled human experiments, and surveys. 
    We observe that all previous studies on pull request in GitHub use either qualitative or quantitative methods, while only one mixed approach combining both with a very time consuming manual analysis for the qualitative part~\cite{kalliamvakou2014promises}, which is quite different from ours, because we apply crowdsourcing directly on the results extracted from primary qualitative studies in a relatively much smaller  time cost.

            \begin{table*}[t!]
\centering
\caption{Features Used in Related Works. $\square$ indicates whether the feature is used or not, while $\blacksquare$ indicated the features are found to be heavily related to the results of pull requests in the according paper.}

  \label{all_features}
\begin{adjustbox}{max width=6.5in}
\begin{tabular}{lllccccccc}
\textbf{Category}      & \textbf{Fetures}   & \textbf{Description}                                                                & \textbf{~\cite{gousios2014dataset}} & \textbf{~\cite{gousios2014exploratory}} & \textbf{~\cite{tsay2014influence}} & \textbf{~\cite{yu2015wait}} & \textbf{~\cite{zhang2014investigating}} & \textbf{~\cite{rahman2014insight}} & \textbf{Ours}  \\ \hline
\multicolumn{1}{l|}{Pull Request} & lifetime\_minites             & Minutes between opening and closing                                                            & $\square$                                  &                                                &                                           &                                    &                                                &                                           &                \\
\multicolumn{1}{l|}{Pull Request} & mergetime\_minutes            & Minutes between opening and merging (only for merged pull requests)                            & $\square$                                  &                                                &                                           &                                    &                                                &                                           &                \\
\multicolumn{1}{l|}{Pull Request} & num\_commits                  & Number of commits                                                                              & $\square$                                  & $\square$                                      & $\square$                                 & $\blacksquare$                     &                                                &                                           & $\square$      \\
\multicolumn{1}{l|}{Pull Request} & src\_churn                    & Number of lines changed (added + deleted)                                                      & $\square$                                  & $\blacksquare$                                 &                                           & $\blacksquare$                     &                                                &                                           & $\square$      \\
\multicolumn{1}{l|}{Pull Request} & test\_churn                   & Number of test lines changed                                                                   & $\square$                                  & $\square$                                      &                                           &                                    &                                                &                                           &                \\
\multicolumn{1}{l|}{Pull Request} & files\_added                  & Number of files added                                                                          & $\square$                                  &                                                &                                           &                                    &                                                &                                           &                \\
\multicolumn{1}{l|}{Pull Request} & files\_deleted                & Number of files deleted                                                                        & $\square$                                  &                                                &                                           &                                    &                                                &                                           &                \\
\multicolumn{1}{l|}{Pull Request} & files\_modified               & Number of files modified                                                                       & $\square$                                  &                                                &                                           &                                    &                                                &                                           &                \\
\multicolumn{1}{l|}{Pull Request} & files\_changed                & Number of files touched (sum of the above)                                                     & $\square$                                  & $\square$                                      & $\square$                                 &                                    &                                                &                                           &                \\
\multicolumn{1}{l|}{Pull Request} & src\_files                    & Number of source code files touched by the pull request                                        & $\square$                                  &                                                &                                           &                                    &                                                &                                           &                \\
\multicolumn{1}{l|}{Pull Request} & doc\_files                    & Number of documentation (markup) files touched                                                 & $\square$                                  &                                                &                                           &                                    &                                                &                                           &                \\
\multicolumn{1}{l|}{Pull Request} & other\_files                  & Number of non-source, non-documentation files touched                                          & $\square$                                  &                                                &                                           &                                    &                                                &                                           &                \\
\multicolumn{1}{l|}{Pull Request} & num\_commit\_comments         & The total number of code review comments                                                       & $\square$                                  &                                                &                                           &                                    &                                                &                                           &                \\
\multicolumn{1}{l|}{Pull Request} & num\_issue\_comments          & The total number of discussion comments                                                        & $\square$                                  &                                                &                                           &                                    &                                                &                                           &                \\
\multicolumn{1}{l|}{Pull Request} & num\_comments                 & The total number of comments (discussion and code review)                                      & $\square$                                  & $\square$                                      & $\blacksquare$                            & $\blacksquare$                     &                                                &                                           & $\square$      \\
\multicolumn{1}{l|}{Pull Request} & num\_participants             & Number of participants in the discussion                                                       & $\square$                                  & $\square$                                      &                                           &                                    &                                                &                                           &                \\
\multicolumn{1}{l|}{Pull Request} & test\_inclusion               & Whether or not the pull request included test cases                                            &                                            &                                                & $\square$                                 & $\square$                          &                                                &                                           &                \\
\multicolumn{1}{l|}{Pull Request} & prior\_interaction            & The number of events the submitter has participated in this project before this pull request   &                                            &                                                & $\square$                                 &                                    &                                                &                                           &                \\
\multicolumn{1}{l|}{Pull Request} & social\_distance              & Whether or not the submitter follows the user who closes the pull request                      &                                            &                                                & $\blacksquare$                            & $\square$                          &                                                &                                           & $\square$      \\
\multicolumn{1}{l|}{Pull Request} & strength of social connection & The fraction of team members that interacted with the submitter in the last three months       &                                            &                                                &                                           & $\square$                          &                                                &                                           &                \\
\multicolumn{1}{l|}{Pull Request} & description complexity        & Total number of words in the pull request title and description                                &                                            &                                                &                                           & $\square$                          &                                                &                                           &                \\
\multicolumn{1}{l|}{Pull Request} & first human response          & Time interval in minutes from pull request creation to first response by reviewers             &                                            &                                                &                                           & $\blacksquare$                     &                                                &                                           & $\square$      \\
\multicolumn{1}{l|}{Pull Request} & total CI latency:             & Time interval in minutes from pull request creation to the last commit tested by CI            &                                            &                                                &                                           & $\blacksquare$                     &                                                &                                           & $\square$      \\
\multicolumn{1}{l|}{Pull Request} & CI result:                    & Presence of errors and test failures while running Travis-CI                                   &                                            &                                                &                                           & $\blacksquare$                     &                                                &                                           & $\square$      \\
\multicolumn{1}{l|}{Pull Request} & mention-@                     & Weather there exist an @-mention in the comments                                               &                                            &                                                &                                           &                                    & $\square$                                      &                                           &                \\ \hline
\multicolumn{1}{l|}{Repository}   & sloc                          & Executable lines of code at creation time.                                                     & $\square$                                  & $\blacksquare$                                 &                                           &                                    &                                                &                                           & $\square$      \\
\multicolumn{1}{l|}{Repository}   & team\_size                    & Number of active core team members during the last 3 months prior to creation                  & $\square$                                  & $\blacksquare$                                 & $\square$                                 & $\square$                          &                                                & $\blacksquare$                            & $\square$      \\
\multicolumn{1}{l|}{Repository}   & perc\_external\_contribs      & The ratio of commits from external members over core members in the last 3 months              & $\square$                                  & $\blacksquare$                                 &                                           &                                    &                                                &                                           & $\square$      \\
\multicolumn{1}{l|}{Repository}   & commits\_on\_files\_touched   & Number of total commits on files touched by the pull request in the past 3 months              & $\square$                                  & $\blacksquare$                                 &                                           & $\square$                          &                                                &                                           & $\blacksquare$ \\
\multicolumn{1}{l|}{Repository}   & test\_lines\_per\_kloc        & Executable lines of test code per 1,000 lines of source code                                   & $\square$                                  & $\blacksquare$                                 &                                           &                                    &                                                &                                           &    $\square$            \\
\multicolumn{1}{l|}{Repository}   & test\_cases\_per\_kloc        & Number of test cases per 1,000 lines of source code                                            & $\square$                                  &                                                &                                           &                                    &                                                &                                           &                \\
\multicolumn{1}{l|}{Repository}   & asserts\_per\_kloc            & Number of assert statements per 1,000 lines of source code                                     & $\square$                                  &                                                &                                           &                                    &                                                &                                           &                \\
\multicolumn{1}{l|}{Repository}   & watchers                      & Project watchers (stars) at creation                                                           & $\square$                                  &                                                & $\blacksquare$                            &                                    &                                                &                                           & $\square$      \\
\multicolumn{1}{l|}{Repository}   & repo\_age                     & How long a project has existed on GitHub since the time of data collection                     &                                            &                                                & $\square$                                   & $\square$                          &                                                &                                           &                \\
\multicolumn{1}{l|}{Repository}   & workload                      & Total number of pull requests still open in each project at current pull request creation time &                                            &                                                &                                           & $\square$                          &                                                &                                           &                \\
\multicolumn{1}{l|}{Repository}   & integrator availability       & The minimum number of hours until either of the top 2 integrators are active during 24 hours   &                                            &                                                &                                           & $\square$                          &                                                &                                           &                \\
\multicolumn{1}{l|}{Repository}   & project maturity              & The number of forked projects as an estimate of the maturity of a base project                 &                                            &                                                &                                           &                                    &                                                & $\blacksquare$                            & $\square$      \\ \hline
\multicolumn{1}{l|}{Developer}    & prev\_pullreqs                & Number of pull requests submitted by a specific developer, prior to the examined one           & $\square$                                  & $\blacksquare$                                 &                                           &                                    &                                                &                                           & $\blacksquare$ \\
\multicolumn{1}{l|}{Developer}    & requester\_succ\_rate         & The percentage of the developer’s pull requests got merged before creation of this one         & $\square$                                  & $\blacksquare$                                 &                                           & $\square$                          &                                                &                                           & $\blacksquare$ \\
\multicolumn{1}{l|}{Developer}    & followers                     & Followers to the developer at creation                                                         & $\square$                                  &                                                & $\square$                                 & $\square$                          &                                                &                                           &                \\
\multicolumn{1}{l|}{Developer}    & collaborator\_status          & The user's collaborator status within the project                                              &                                            & $\square$                                      & $\blacksquare$                            & $\square$                          &                                                &                                           & $\square$      \\
\multicolumn{1}{l|}{Developer}    & experience                    & Developers‘ working experience with the project                                                &                                            &                                                &                                           &                                    &                                                & $\blacksquare$                            & $\square$      \\ \hline
\multicolumn{1}{l|}{Other}        & Friday effect                 & True if the pull request arrives Friday                                                        &                                            &                                                &                                           & $\square$                          &                                                &                                           &               
\end{tabular}
\end{adjustbox}

\end{table*}
    Table~\ref{all_features} summarizes the most representative features these studies state are relevant to determining
    the fate of a pull request. Note that different studies found that different features were most relevant
    to deciding what happens to pull requests.  In that table:
    \begin{itemize}
    \item
    White boxes $\square$ denote that a paper examined that feature;
    \item
    Black boxes $\blacksquare$ denote when  that paper concluded that feature was important;
     \end{itemize}
    The last column shows what lessons   we took from these prior studies.
      \begin{itemize}
    \item If any other column marked
    a feature as important, then we added it into the set of features we examined. Such features are denoted
    with a white box $\square$  in the last column.
    \item Later in this paper, we run feature subset selectors on the data to determine which features
    are most informative. Such  features are denoted with a black box $\blacksquare$.
      \end{itemize}

    \subsection{Map Insights into Questions and Features} \label{Task Template}

    The tasks performed by the crowd were designed to collect quantitative information about the pull requests, which could be checked against a ground truth extracted programmtically (e.g., was the pull request accepted?), and also collect information related to the pull request discussion, described next. 
    
    The primary study~\cite{tsay2014let} concluded, 
amongst other things, that:
    \begin{quote}
		\textit{Issues raised around code contributions are mostly disapproval for the problems being solved, disapproval for the solutions and suggestion for alternate solutions.}\\
		
		\noindent \textit{Methods to affect the decision making process for pull requests are mainly by offering support from either external developers or core members.}
	\end{quote}
	In order to use crowdsourcing to do a case study for pull requests, our tasks contained questions related to five concepts. 
	These five concepts reference  important  findings from TDH's work, and are also treated as the assumptions we are going to validate:
	
	\begin{enumerate}
      \item \textit{Is there a comment showing support for this pull request, and from which party?} 
      The crowd's answer
      to this question lets us define three binary variables: Q1\_support, Q1\_spt\_core, Q1\_spt\_other, 
      which stands for ``support showed", ``support from core members" and ``support from other developers" respectively.
      
      \item \textit{Is there a comment proposing alternate solutions, and from which party? }
      The crowd's answer
      to this question lets us define three binary variables:
      Q2\_alternate\_solution, Q2\_alt\_soln\_core, Q2\_alt\_soln\_other, 
      which stands for ``alternate solution proposed", ``alternate solution proposed by core members" and ``alternate solution proposed by other developers" respectively.
      
      \item \textit{Did anyone disapprove the proposed solution in this pull request, and for what reason?}
      The crowd's answer
      to this question lets us define four binary variables: Q3\_dis\_solution, Q3\_dis\_soln\_bug, Q3\_dis\_soln\_improve, Q3\_dis\_soln\_consistency, 
      which stands for ``disapproval for the solution proposed", ``disapproval due to bug", ``disapproval because code could be improved" and "disapproval due to consistency issues" respectively.
      
      \item \textit{Did anyone disapprove the problems being solved? E.g question the value or appropriateness of this pull request for its repository.}
      The crowd's answer
      to this question lets us define three binary variables:
      Q4\_dis\_problem, \newline
      Q4\_dis\_prob\_no\_value, \newline 
      Q4\_dis\_prob\_not\_fit,
      which stands for ``disapproval for the problem being solved", ``disapproval due to no value for solving this problem" and ``disapproval because the problem being solved does not fit the project well" respectively.

      \item \textit{Does this pull request get merged/accepted?}
      The crowd's answer to this question lets us define a class variable for this system.
    \end{enumerate}
    The full version of our questions are   available on-line\footnote{\url{http://dichen.me/fse17/mt_template.html}}.
    To those questions we also added three preliminary questions that require crowd workers to identify the submitter, core members and external developers for each pull request. 
    These extra questions served
	 two purposes: First, they let a crowd worker grow familiar with analyzing pull request discussions. Second, they let us reject
	 answers from unqualified crowd workers since we could programatically  extract the ground truth from the repository for comparison. 

    We also extract answers for these questions from the results in TDH. These answers are served as ``gold" standard tasks which enable quality control during crowdsourcing and sanity checking for the answers after crowdsourcing.

    \
    \subsection{Data Expansion and Cost Control}
    \label{pr selection}
    
    To make sure the pull requests are statistically similar to those of TDH's work~\cite{tsay2014let, tsay2014influence}, we applied similar selection rules on 612,207 pull requests that were opened during January 2016 from GHTorrent~\cite{Gousi13GHTorrent}, which is a scalable, searchable, offline mirror of data offered through the GitHub Application Programmer Interface (API). The selection criteria are stated as follows:
    \begin{enumerate}
      \item Pull requests should be closed (558,480 left).
      \item Pull requests should have comments (50,440 left, with 2, 3, and 7 comments as the 25, 50, 75 percentiles, respectively. 
      \item Pull request comment number should be above 8.
      \item Exclude pull requests whose repository are forks to avoid counting the same contribution multiple times.
      \item Exclude pull requests whose last update is late than January, 2016, so that we can make sure the project is still active (8,438 left).      
      \item Retain only pull requests with at least 3 participants and where the repository has at least 10 forks and 10 stars (565 left).
    \end{enumerate}
    From these 565 pull requests, we sampled 210 such that half were ultimately merged and the other half were rejected.

    
   The  210 pull requests were
    published on the Amazon Mechanical Turk (MTurk) crowdsourcing platform for analyzing in 2 rounds, together with the 20 carefully studied pull requests from TDH~\cite{tsay2014let} inserted for each round as ``gold" standard tasks. The 1st round has 100(80+20) pull requests in total, while 2nd round has 150(130+20) pull requests in total. 
    
    \paragraph{{\bf Cost Control}} \label{Cost Control}
    We want to make sure the cost is as low as possible but also provide a fair payment for the participants. 
    According to several recent surveys on MTurk~\cite{buhrmester2011amazon, berinsky2012evaluating, paolacci2010running, ipeirotis2010demographics}, the average hourly wage is \$1.66 and MTurk workers are willing to work at \$1.40/hour. We estimated about 10 minutes needed for each HIT, and first launched our task with \$0.25 per HIT but only received 1 invalid feedback after 2 days. So we doubled our payment to \$0.50 for each HIT, which requires to analyze one single pull request.  Each round of tasks were completed in one week. Our final results show that 17 minutes are spent for each HIT on average, which means \$1.76 per hour. In total, 27 workers participated in our tasks, and 77 hours of crowd time were spent to get all the pull requests studied.
    
    \subsection{Crowdsourcing Quality Control} \label{quality control in cs}
    A major issue in crowdsourcing
    is how to reduce the noise inherent
    in data collection 
    from such a subjective source
    of information. This section
    describes the three operators
    we used to increase data quality:
    \begin{itemize}
        \item  Audience screening;
        \item ``Gold" standard questions and tasks;
        \item Feature subset selection.
    \end{itemize}
    Quality and experience filters were applied to screen potential participants; only workers with HIT approval rate above 90\%, and who had completed at least 100 approved HITs could participate. 
    
    Next, a domain-specific screening process was applied. To make sure the crowd participants are qualified to analyze the pull requests in our study, we require them to be GitHub users and answer preliminary questions related to identifying the pull request key players and pull request acceptance {\em on every pull request analyzed}. These are questions for which we can systematically extract values from the pull requests; if these golden questions are answered incorrectly, the task was rejected and made available to other crowd workers.  
    
    \paragraph{{\bf Quality Control}} \label{Quality Control}
    \vspace{-0.25cm}
    Part of the audience selection relates to quality control since the workers were required to have demonstrated high quality in prior tasks on the MTurk platform and answer simple questions about the pull request correctly. Beyond that, we used (1) redundant question formats~\cite{stolee2015exploring} and (2) gold standard {\em tasks} to control crowd quality. 
    
    
    (1) For each question in the task related to the pull request comment discussion, we require workers to answer a yes/no question and then copy the comments supporting their answers from the pull request into the text area under each question. Take question 1 for example (\textit{Is there a comment proposing alternate solutions?}): if they choose \textit{"Yes, from core members"}, then they need to copy the comments within the pull requests to the text area we provided. 
    
    (2) This study was run in two phases, one phase with 80 out of the 210 new pull requests, and one phase with the remaining 130. In each phase, the original 20 pull requests were added to the group. The tasks were randomly assigned to crowd workers. For those crowd workers who got one of the 20 previously studied pull requests, we checked their answers against the ground truth~\cite{tsay2014let}; inaccurate responses were rejected and those workers were blocked. This acted as a random quality control mechanism.  
    
    In total, we have 250 highly discussed pull requests from 142 projects and analyzed by 27 workers. After filtering out the unqualified ones using the control processes stated above, 190 pull requests were left with 3,471 comments. The unqualified responses were a result of part (1) above, but an operational error led us to approve the tasks despite the poor comment quality, leaving us with a smaller data set for further analysis.
    
    For the rest of this paper, we will refer to the data collected via crowd 
    as the {\em qualitative} pull request features.
    
    \subsection{Quantitative Analysis} \label{feature_selection}
    
    In this study,  we have 2 groups of features\: (1)~all the quantitative features found important in previous works and (2)~all qualitative features extracted from the results of studying pull requests in detail by the qualified crowd. For each group of features, we run the CFS feature selector~\cite{hall1999correlation} to reduce the features to use for our decision tree classifier.    
  To collect the quantitative features, we started with Table \ref{all_features} and used the GitHub API
  to extract the features marked in the   right-hand-side column.

    CFS evaluates and ranks feature subsets.
     One reason to use CFS over, say, correlation, is that CFS
   returns {\em sets} of useful features while simpler
   feature selectors do not  understand the interaction between
   features.
   
   CFS assumes that a ``good'' set of features   contains features that are highly connected with the target class, but weakly connected to each other. 
    To implement this heuristic,
    each feature subset is scored as follows according to Hall et al. ~\cite{hall1999correlation}:
    \[
    \mathit{merit}s = \frac{k\overbar{r_{\mathit{cf}}}}{ \sqrt{k+k(k-1)\overbar{r_{\mathit{ff}}}}}
    \]
    where 
    $\mathit{merit}s$ is the value of some subset $s$ of the
    features containing $k$ features;
    $r_{\mathit{cf}}$ is a score describing the connection of that feature
    set to the class;
    and $r_{\mathit{ff}}$ is the mean score of the feature to feature
    connection between the items in $s$.
    Note that for this fraction to be maximal, $r_{\mathit{cf}}$ must be large
    and $r_{\mathit{ff}}$ must be small, which means features have to correlate
    more to the class than each other.
    
    This equation is used to guide a  best-first search with a horizon of five  to select most informative set of features. 
    Such a search proceeds as  follows. The initial frontier is all sets containing one different feature. The frontier of size $n$, which initialized with $1$, is sorted according to {\em merit} and the best item is grown to all sets of size {\em n+1} containing the best item from the last frontier. The seaerch
    stops when no improvement have been seen in last five frontiers in $merit$. Return the best subset seen so far when stop.



    
    Our experiments assessed
three groups of features:
\begin{enumerate}
\item
    After CFS feature selector, the selected {\em quantitative features} were 
    \textit{commits\_on\_files\_touched, 
    requester\_succ\_rate, 
    prev\_pullreqs}, 
    which are quite intuitive. 
    \item
    
    The second group
    of   {\em crowdsourced features} were
    \textit{Q3\_dis\_s, \newline
    Q4\_dis\_p\_nv};
    \item 
    The third group of {\em combined
    features} were   the 
    combination of both quantitative and crowdsourced features; i.e. 
    \textit{ Q3\_dis\_s, 
    commits\_on\_files\_touched, 
    requester\_succ\_rate, \newline
    prev\_pullreqs}.
\end{enumerate} 
For each of these three sets of features, we  ran a 10x5 cross validation for supervised learning with the 3 different groups of features. 
These generate three models that predicted if a pull request would get merged/accepted or not. 
A decision tree learner was used as our supervised learning algorithm.  This was selected
after our initial studies with several other learners that proved to
be less effective in this domain (Naive Bayes and SVM).

\section{Results}

 Using Amazon's Mechanical Turk micro-task crowdsourcing platform, we collect data for 1) the original 20 pull requests from the primary study~\cite{tsay2014let}, and 2) 210 additional, independent pull requests, an order of magnitude more than pull requests than the primary study. 
    This data includes qualitative information about the pull request discussion, such as whether there is a comment showing support, proposing an alternate solution, disapproving of the solution, and disapproving of the problem being solved.
    The benefits of the larger sample size is two-fold. 
    First, by using similar selection criteria in the secondary study compared to the primary study, we are able to check the 
    stability and external validity of the findings in the primary study using a much larger sample (RQ1, RQ2). 
    Second, in terms for informativeness, we can extract features from crowd's answers, 
    which is qualitative, and build models to predict pull request acceptance results. 
    This allows us to compare the performance of models built with (a)~the features identified as important in the primary study and (b)~the features from related, quantitative works (RQ3).
    
    \subsection{RQ1: Can the crowd reproduce prior results quickly and cheaply?} \label{RQ_1}

\begin{table}[!t]
\centering
\begin{adjustbox}{max width=2in}
\begin{tabular}{|l|l|l|l|}
\hline
\rowcolor[HTML]{C0C0C0} 
\textbf{Questions} & \textbf{Precistion} & \textbf{Recall} & \textbf{F1-Score} \\ \hline
\textbf{Q1}        & 0.769               & 0.769           & 0.770             \\ \hline
\textbf{Q2}        & 0.818               & 0.750           & 0.783             \\ \hline
\textbf{Q3}        & 0.727               & 0.667           & 0.696             \\ \hline
\textbf{Q4}        & 0.778               & 0.700           & 0.737             \\ \hline
\textbf{Q5}        & 0.833               & 0.714           & 0.770             \\ \hline
\textbf{Total}     & \textbf{0.801}      & \textbf{0.742}  & \textbf{0.770}    \\ \hline
\end{tabular}
\end{adjustbox}
\caption{Quality for Crowdsourcing Results from Amazon Mechanical Turk (RQ1).}
\label{mt_quality}
\vspace{-0.46cm}
\end{table}


   RQ1
   checks if our analysis in {\S}~\ref{Task Template} correctly captured the essence of the TDH study.
   
   In this test, we used the 20 ``gold'' task results from TDH. Each pull request was labeled with the 
   gold results. Next, we checked the performance  of the  crowd with respect
   to the gold results.  
   
    Table \ref{mt_quality} shows the   $precision$, $recall$ and $F_1$ scores of the crowd working
    on the gold tasks. As    seen in Table~\ref{mt_quality}, the precision and recall of these `human predictors' on the 20 gold tasks is 80\% and 74\% respectively (so  $F_1 \approx 77\%$). Based on our prior
    work with data mining from SE data, we assert that these values
    represent a close correspondence between the TDH results and those from the crowd.
    
    
    
    To make sure we did not  mistakenly analyze the crowd's answers, we hand-examined the cases where the crowd disagreed with the TDH. Interestingly, we found several cases that crowd workers appear to be correct. For example, TDH classify the 17th pull request they studied as no support, while the crowd found the comment from the user {\tt drohthlis} saying `This is great news!', which is an apparent indicator for the supporting this pull request after our examination. 
    Another two cases are the 16th and 20th pull requests they studied. Crowd workers found  clear suggestions for alternative solutions (i.e., `What might be better is to ...', `No, I think you can just push -f after squashing.'), which TDH does not find.

As to the issue
of speed and cost, 
    TDH report that they required about 47 hours to collect interview
     data on 47 users within which, they investigated the practices about pull requests. TDH does not report the subsequent analysis time but, given the qualitative nature of their methods,
     we conjecture that took hours to weeks.  
     
     By way of comparison, 
   we spent \$200, to buy 77  hours of crowd time. In that time,   250 pull requests were analyzed (100, 150 pull requests respectively for each round). Note that, in this study, we  included the 20 pull requests already studied by TDH. 
   
In summary, we answer RQ1 in the affirmative.

	 \subsection{RQ2: Are the primary study's results stable? } \label{RQ_2}

    \begin{figure}[!t]
        \centering
        {\includegraphics[width=3in]{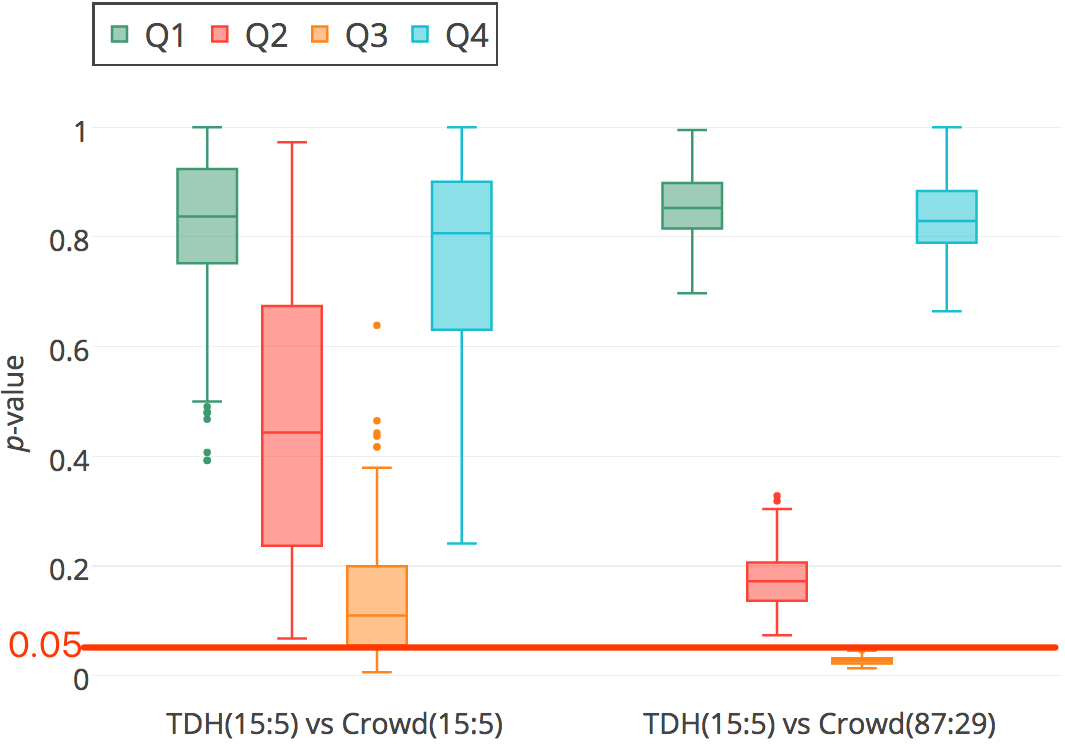}}
	    \caption{Stability Checking: \textit{p}-values for comparing unscaled and scaled crowd answers with answers from TDH (Interactive version also available at \url{http://tiny.cc/mosss-1})}
		\label{fig:stability_box}	
    \end{figure}     
    
     As described in the introduction, one motivation for this work was
     checking if crowdsourcing can scale and confirm the  external validity of
     qualitative conclusions. 
     This issue is of particular concern for crowdsoucing studies due the subjective nature
     of the opinions from the crowd. If those opinions {\em increased} the variance of the collected
     data, then the more data we collect, the {\em less} reliable the conclusions.
     
     To test for this concern, we compare the pull requests studied by crowd (excluding the 20 gold tasks) with the 20 pull request studied by TDH (15 merged, 5 rejected). We first randomly select 15 merged and 5 rejected pull requests studied by crowd  100 times, so that we can compare the these 2 independent samples at the same scale and with the same distribution. Then we run another 100 iteration for randomly selecting 87 merged and 29 rejected pull requests studied by crowd, which still has the same distribution but at a 6 times larger scale. \textit{p}-values are collected for each sample comparison in the 2 runs.
     
     Figure~\ref{fig:stability_box} shows the results of comparing pull requests from TDH and an independent sample with 2 different scales. As shown, Questions 1, 2, 4 are quite stable for both scales. Moreover, Question 1 and 4 are becoming more stable when scale becomes larger, while Question 2 becomes less stable at a larger scale. For Question 3, all of the p-values are lower than 0.05 at the large scale, though the median of its p-value is higher than 0.05 at the same scale as TDH. This may indicate that TDH did not cover enough pull requests to achieve a representative sample for the finding, which is mapped into Question 3 about disapproving comments.

     Accordingly, we answer RQ2 in the affirmative. The results for Q1, Q2, and Q4 do not differ significantly between TDH and independent samples of the same size or of a larger size. The exception is Q3, for which the results differ significantly when scaling to a larger data set.


    
    
     
    \subsection{RQ3: How well can the qualitative and quantitative features predict PR acceptance?} \label{RQ_3}


    \begin{figure}[!t]
        \centering
        {\includegraphics[width=3in]{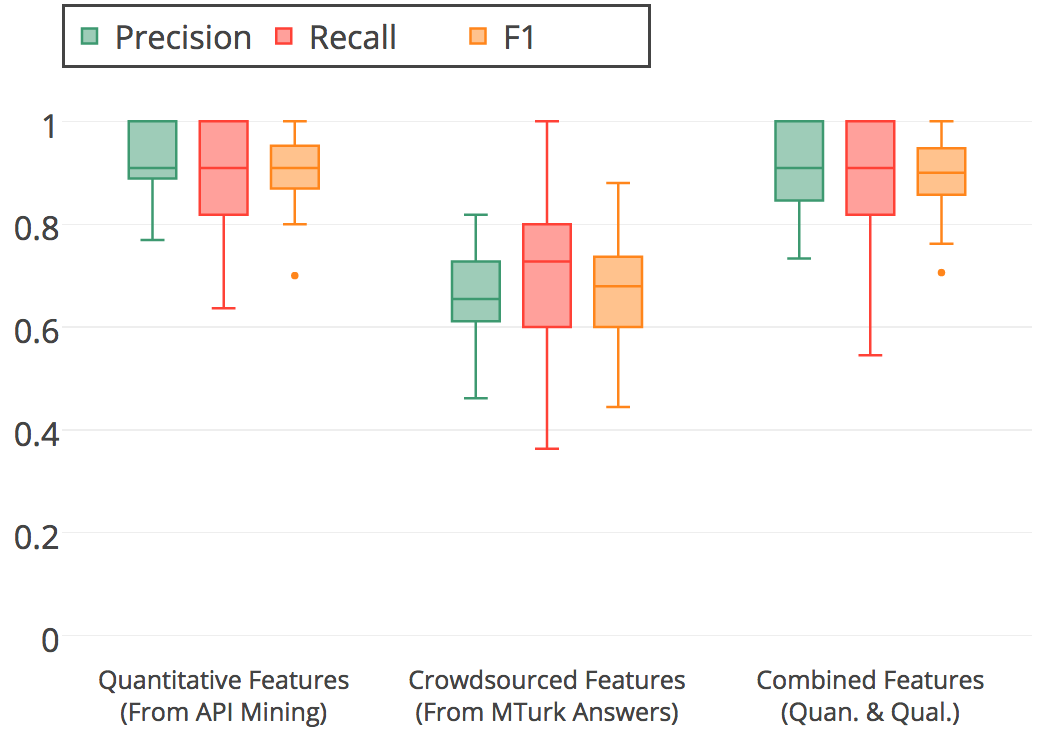}}
	    \caption{Performance Comparison for Using different feature set to predict whether a pull requests will be accepted. (Interactive version also available at \url{http://tiny.cc/mosss-2})}
		\label{fig:performance}
    \end{figure}
    
    The results are shown in Figure \ref{fig:performance}, expressed
    in terms of precision, recall, and the $F_1$
    score; i.e.
    the harmonic mean
    of precision and 
    recall, for each of three feature sets: quantitative, crowdsourced, and a combination. 
    Note that  the
    performances of the predictor using crowdsourced features 
    are not as high or as stable as the one built with quantitative features. 
    We can see that:
    \begin{itemize}
        \item The selected quantitative features achieved $F_1$ score at 90\% with a range of 20\%;
        \item The selected crowdsourced features achieved lower $F_1$ score at 68\% with a larger range.
        \item The combined selected features did better than just using quantitative; but performed no better than just using only the quantitative features.
    \end{itemize}.

At first glance, the models learned from crowdsourced features
   performed worse than using  quantitative features extracted from numerous prior
   data mining studies.  But is this the case?
   Do the middle results really reflect the insights taken from \fig{performance}? As seen in {\S}~\ref{RQ_1}, we have shown that our analysis in {\S}~\ref{Task Template} correctly captured the essence of the TDH study.
    
    In summary, we answer RQ3 in the affirmative:  
    \begin{enumerate}
    \item The middle results of Figure~\ref{fig:performance} adequately reflect the TDH results;
    \item Accordingly, we can also conclude  that Figure~\ref{fig:performance} shows     the TDH results 
    being out-performed by the purely quantitative
    features.
    \end{enumerate}

     \begin{table*}[t!]
\small
\caption{Comparison of Different Methods for GitHub Pull Requests Studies. '$\surd$' stands for True, '$\times$' for False and '?' for Uncertain.}
\begin{adjustbox}{width=5in}
\begin{tabular}{|l|l|l|l|}
\hline
\textbf{Measures}        & \cellcolor[HTML]{C0C0C0}\textbf{\begin{tabular}[c]{@{}l@{}}Quantitative Studies: \\ E.g. API-Mining, Data Modeling\end{tabular}} & \cellcolor[HTML]{656565}{\color[HTML]{EFEFEF} \textbf{\begin{tabular}[c]{@{}l@{}}Qualitative Studies: \\ E.g. Interview, Case Study\end{tabular}}} & \cellcolor[HTML]{343434}{\color[HTML]{FFFFFF} \textbf{\begin{tabular}[c]{@{}l@{}}Crowdsourcing Assisted \\ Qualitative Studies\end{tabular}}} \\ \hline
\textbf{Time}            & Hours                                                                                                                            & Weeks or Months                                                                                                                                    & Days                                                                                                                                          \\ \hline
\textbf{Cost}            & \$10-100                                                                                                                      & \$1,000-3,000                                                                                                                                       & \$100-300                                                                                                                                     \\ \hline
\textbf{Perspective}     & Objective                                                                                                                        & Subjective                                                                                                                                         & Combined                                                                   \\ \hline
\textbf{Purpose}         & \begin{tabular}[c]{@{}l@{}}Descriptive: Reveal General Trends\\ Explanatory: Summarize Patterns\end{tabular}                     & \begin{tabular}[c]{@{}l@{}}Exploratory: Find Details\\ Improving: Study Drawbacks\end{tabular}                                                     & \begin{tabular}[c]{@{}l@{}}Confirmatory: Theories\\ Scale Up \& Lower Costs\end{tabular}                                                      \\ \hline
\textbf{Stability}       & \textbf{$\surd$}                                                                                                                    & \textbf{?}                                                                                                                                  & \textbf{$\surd$}   
 \\ \hline
\textbf{Scalability}     & \textbf{$\surd$}                                                                                                                    & \textbf{$\times$}                                                                                                                                     & \textbf{$\surd$}    
\\ \hline
\textbf{Reproducibility} & \textbf{$\surd$}                                                                                                                    & \textbf{$\times$}                                                                                                                                     & \textbf{$\surd$}                                                                                                                                         
\\ \hline
\textbf{Can Build Predictor}     & \textbf{$\surd$}                                                                                                                           & \textbf{$\times$}                                                                                                                                            & \textbf{$\surd$}                                                                                                 
\\ \hline
\end{tabular}
\end{adjustbox}
\centering

\label{method_comparison}
\end{table*}

\section{Threats to Validity}
	As with any empirical study, biases can affect the final
    results. Therefore, any conclusions made from this work must
    be considered with the following issues in mind:

 \textbf{Sampling bias:} This threatens any classification experiment;
        i.e., what matters there may not be true here. For example,
        the pull requests used here are selected using the rules described in \ref{pr selection}. Only 250 highly discussed pull requests from active projects are sampled and analyzed, so our results may not reflect the patterns for all the pull requests. That said, we note that one reason to endorse crowdsourcing is that its sample size can be orders of magnitude larger than using just qualitative methods.  For example, TDH reported results from just 20 pull requests.
        
  \textbf{Learner bias:} For building the acceptation predictors in this
        study, we elected to use a decision tree classifier. We chose   decision trees
        because it suits for small data samples and its results were comparable to 
        the more complicated algorithms like Random Forest and SVM. Classification is a large and active field
        and any single study can only use a small subset of the known
        classification algorithms. Future work should repeat this study using other learners.
        
   \textbf{Evaluation bias:} This paper uses \textit{precision}, \textit{recall} and  $F_1$ \textit{score} measures   of predictor's performance. Other performance measures used in software engineering
        include accuracy and entropy.  Future work should repeat this study using different evaluation biases.
      
     \textbf{Order bias:} For the performance evaluation part, the order that the data trained  and predicted affects the results. To mitigate this order bias, 
      we run the 5-bin cross validation 10 times randomly changing the order of the pull requests each time.
        
        

  \section{Discussion}   
    We summarize our experience with qualitative, 
    quantitative and crowdsourcing methods in Table \ref{method_comparison}. 
    As shown, the outstanding benefit of quantitative methods is 
        to find general patterns or underlying trends. They usually work with large amounts of data and are fast to deploy, cheap to run, and easy to replicate. 
    The major drawback is they often ignore the details and often lack a human-level understanding.
        Besides, it's also hard to implement a data miner for large projects,
        and time-consuming for complex systems.
        
 
    The outstanding benefit of qualitative  methods is 
        their exploratory nature to generate new theorems or improve existing ones. 
        They seek details  to understand what  humans really care about and find insights. 
    While the major drawback is the sample size is restricted to a very small size compared to quantitative method, because of the involvement of human. Therefore, the complexity, time and/or
        money cost are needed to be taken into consideration. 
    Further, most of the studies with qualitative methods are 
        hard to replicate or scale up for larger sample size.

    The outstanding benefit of crowdsourcing  methods is 
            they try to mix and match methods in order to exploit the strengths of all the above approaches. 
            They could offer the human-level understanding or intelligence compared with quantitative methods and also scale up and increase the diversity of data size due to its low costs and massive work force compared with qualitative methods.
        The major concerns of crowdsourcing methods are 
        that  the quality and knowledge of the workers from crowd are inadequate, and that the study context cannot be as controlled.  
        Following this,  it is unclear if the crowd can serve as a proxy for domain experts.
           Also, it is important to ask about the right questions for the crowd to get the expected results. 

     
        


    The key point of this paper is that 
    it is misleading to review the benefits and drawbacks of these methods
    {\em in isolation}. Rather, it is more insightful
    to consider what these methods can achieve {\em in combination}. 
                For example, in the original qualitative TDH study, the authors found that 1)~\textit{supports}, 2)~\textit{alternate solutions}, 3)~\textit{disapproval for the proposed solutions}, and 4)~\textit{disapproval for the problems being solved} were important factors that guard pull requests' acceptance. In this scaled, crowdsourced replication, we found that factors 1, 2 and 4 still hold, but 3 was unstable. Thus, this combination of empirical methods allows us to pinpoint more precisely results that are steadfast against tests of external validity and the results that need further investigation.
                
    In the end, the secondary quantitative study would have been impossible without the primary qualitative work, and we should make best use of the time-consuming qualitative works, instead of stopping after we get results from qualitative results (and vice versa).
    We find qualitative studies can inspire quantitative studies by care-fully mapping out areas of concern.  Primary  qualitative  study  can also  provide  the  data  needed  to control secondary quantitative crowdsourcing studies. We also find a single  primary  qualitative  study  can  direct  the  work  of many secondary quantitative studies, and our work is just one example of the secondary studies after TDH's qualitative work.

\section{Conclusion}

     In this paper, we designed MOSSS and applied it to one state of art qualitative study from TDH. 
     As seen in Table~\ref{tab:lit_rev} and Table~\ref{all_features}, we reviewed and summarized all related paper on GitHub pull requests from both Google Scholar from 2008 to 2016 and 10 top SE conferences from 1992 to 2016. 
     In {\S}\ref{Task Template} and \ref{pr selection}, we show that, with results and data from TDH's primary study, it is possible to quickly map their insights into micro questions for crowd workers and expand the data they studied to a larger scale. 
     Moreover, from the 20 pull requests studied in TDH, we also extracted answers treated as the ground truth for our questions described in {\S}\ref{Task Template}. 
     These answers served as gold tasks for quality control during our crowdsourcing process in {\S}\ref{quality control in cs}. 
     With these gold tasks, we not only checked the sanity of the qualified answers from crowd in {\S}\ref{RQ_1}, but also checked the stability of the primary results from TDH in {\S}\ref{RQ_2}.
     As seen in Table~\ref{mt_quality} and Figure~\ref{fig:stability_box}, we showed an overall 77\% of $F1$ score and three out of the four findings we extracted from TDH are stable from a larger scale.
     In {\S}\ref{feature_selection}, we did quantitative analysis by applying data mining techniques on the large amount of answers we collected from crowd and build predictors with crowdsourced features, quantitative features from literature review and the combination of both. 
     As shown in Figure~\ref{fig:performance}, we found the crowdsourced features mapped from THD results could do a good job predicting the fate for pull requests, but cannot compete with features selected from related quantitative studies.
     These results have implications for the value of combining diverse empirical methods and 
     for conducting conceptual replications of empirical software engineering studies in new contexts. 

\bibliographystyle{ACM-Reference-Format}
\balance

\begin{thebibliography}{00}


\ifx \showCODEN    \undefined \def \showCODEN     #1{\unskip}     \fi
\ifx \showDOI      \undefined \def \showDOI       #1{{\tt DOI:}\penalty0{#1}\ }
  \fi
\ifx \showISBNx    \undefined \def \showISBNx     #1{\unskip}     \fi
\ifx \showISBNxiii \undefined \def \showISBNxiii  #1{\unskip}     \fi
\ifx \showISSN     \undefined \def \showISSN      #1{\unskip}     \fi
\ifx \showLCCN     \undefined \def \showLCCN      #1{\unskip}     \fi
\ifx \shownote     \undefined \def \shownote      #1{#1}          \fi
\ifx \showarticletitle \undefined \def \showarticletitle #1{#1}   \fi
\ifx \showURL      \undefined \def \showURL       #1{#1}          \fi
\providecommand\bibfield[2]{#2}
\providecommand\bibinfo[2]{#2}
\providecommand\natexlab[1]{#1}
\providecommand\showeprint[2][]{arXiv:#2}

\bibitem[\protect\citeauthoryear{??}{shu}{2013}]%
        {shull13}
 \bibinfo{year}{2013}\natexlab{}.
\newblock \bibinfo{title}{Emprical Software Engineering v2.0}.
\newblock   (\bibinfo{year}{2013}).
\newblock


\bibitem[\protect\citeauthoryear{Amir, Shahar, Gal, and Ilani}{Amir
  et~al\mbox{.}}{2013}]%
        {amir2013verification}
\bibfield{author}{\bibinfo{person}{Ofra Amir}, \bibinfo{person}{Yuval Shahar},
  \bibinfo{person}{Ya'akov Gal}, {and} \bibinfo{person}{Litan Ilani}.}
  \bibinfo{year}{2013}\natexlab{}.
\newblock \showarticletitle{On the verification complexity of group
  decision-making tasks}. In \bibinfo{booktitle}{{\em First AAAI Conference on
  Human Computation and Crowdsourcing}}.
\newblock


\bibitem[\protect\citeauthoryear{Bacchelli and Bird}{Bacchelli and
  Bird}{2013}]%
        {bacchelli2013expectations}
\bibfield{author}{\bibinfo{person}{Alberto Bacchelli} {and}
  \bibinfo{person}{Christian Bird}.} \bibinfo{year}{2013}\natexlab{}.
\newblock \showarticletitle{Expectations, outcomes, and challenges of modern
  code review}. In \bibinfo{booktitle}{{\em Proceedings of the 2013
  international conference on software engineering}}. IEEE Press,
  \bibinfo{pages}{712--721}.
\newblock


\bibitem[\protect\citeauthoryear{Barowy, Curtsinger, Berger, and
  McGregor}{Barowy et~al\mbox{.}}{2012}]%
        {Barowy:2012:API:2398857.2384663}
\bibfield{author}{\bibinfo{person}{Daniel~W. Barowy}, \bibinfo{person}{Charlie
  Curtsinger}, \bibinfo{person}{Emery~D. Berger}, {and} \bibinfo{person}{Andrew
  McGregor}.} \bibinfo{year}{2012}\natexlab{}.
\newblock \showarticletitle{AutoMan: A Platform for Integrating Human-based and
  Digital Computation}.
\newblock \bibinfo{journal}{{\em SIGPLAN Not.\/}} \bibinfo{volume}{47},
  \bibinfo{number}{10} (\bibinfo{date}{Oct.} \bibinfo{year}{2012}),
  \bibinfo{pages}{639--654}.
\newblock
\showISSN{0362-1340}
\showDOI{%
\url{http://dx.doi.org/10.1145/2398857.2384663}}


\bibitem[\protect\citeauthoryear{Begel, Bosch, and Storey}{Begel
  et~al\mbox{.}}{2013}]%
        {begel2013social}
\bibfield{author}{\bibinfo{person}{Andrew Begel}, \bibinfo{person}{Jan Bosch},
  {and} \bibinfo{person}{Margaret-Anne Storey}.}
  \bibinfo{year}{2013}\natexlab{}.
\newblock \showarticletitle{Social networking meets software development:
  Perspectives from github, msdn, stack exchange, and topcoder}.
\newblock \bibinfo{journal}{{\em IEEE Software\/}} \bibinfo{volume}{30},
  \bibinfo{number}{1} (\bibinfo{year}{2013}), \bibinfo{pages}{52--66}.
\newblock


\bibitem[\protect\citeauthoryear{Begel and Zimmermann}{Begel and
  Zimmermann}{2014}]%
        {begel2014analyze}
\bibfield{author}{\bibinfo{person}{Andrew Begel} {and} \bibinfo{person}{Thomas
  Zimmermann}.} \bibinfo{year}{2014}\natexlab{}.
\newblock \showarticletitle{Analyze this! 145 questions for data scientists in
  software engineering}. In \bibinfo{booktitle}{{\em Proceedings of the 36th
  International Conference on Software Engineering}}. ACM,
  \bibinfo{pages}{12--23}.
\newblock


\bibitem[\protect\citeauthoryear{Berinsky, Huber, and Lenz}{Berinsky
  et~al\mbox{.}}{2012}]%
        {berinsky2012evaluating}
\bibfield{author}{\bibinfo{person}{Adam~J Berinsky}, \bibinfo{person}{Gregory~A
  Huber}, {and} \bibinfo{person}{Gabriel~S Lenz}.}
  \bibinfo{year}{2012}\natexlab{}.
\newblock \showarticletitle{Evaluating online labor markets for experimental
  research: Amazon. com's Mechanical Turk}.
\newblock \bibinfo{journal}{{\em Political Analysis\/}} \bibinfo{volume}{20},
  \bibinfo{number}{3} (\bibinfo{year}{2012}), \bibinfo{pages}{351--368}.
\newblock


\bibitem[\protect\citeauthoryear{Blincoe, Sheoran, Goggins, Petakovic, and
  Damian}{Blincoe et~al\mbox{.}}{2016}]%
        {blincoe2016understanding}
\bibfield{author}{\bibinfo{person}{Kelly Blincoe}, \bibinfo{person}{Jyoti
  Sheoran}, \bibinfo{person}{Sean Goggins}, \bibinfo{person}{Eva Petakovic},
  {and} \bibinfo{person}{Daniela Damian}.} \bibinfo{year}{2016}\natexlab{}.
\newblock \showarticletitle{Understanding the popular users: Following,
  affiliation influence and leadership on GitHub}.
\newblock \bibinfo{journal}{{\em Information and Software Technology\/}}
  \bibinfo{volume}{70} (\bibinfo{year}{2016}), \bibinfo{pages}{30--39}.
\newblock


\bibitem[\protect\citeauthoryear{Brunet, Murphy, Terra, Figueiredo, and
  Serey}{Brunet et~al\mbox{.}}{2014}]%
        {brunet2014developers}
\bibfield{author}{\bibinfo{person}{Jo{\~a}o Brunet}, \bibinfo{person}{Gail~C
  Murphy}, \bibinfo{person}{Ricardo Terra}, \bibinfo{person}{Jorge Figueiredo},
  {and} \bibinfo{person}{Dalton Serey}.} \bibinfo{year}{2014}\natexlab{}.
\newblock \showarticletitle{Do developers discuss design?}. In
  \bibinfo{booktitle}{{\em Proceedings of the 11th Working Conference on Mining
  Software Repositories}}. ACM, \bibinfo{pages}{340--343}.
\newblock


\bibitem[\protect\citeauthoryear{Buhrmester, Kwang, and Gosling}{Buhrmester
  et~al\mbox{.}}{2011}]%
        {buhrmester2011amazon}
\bibfield{author}{\bibinfo{person}{Michael Buhrmester}, \bibinfo{person}{Tracy
  Kwang}, {and} \bibinfo{person}{Samuel~D Gosling}.}
  \bibinfo{year}{2011}\natexlab{}.
\newblock \showarticletitle{Amazon's Mechanical Turk a new source of
  inexpensive, yet high-quality, data?}
\newblock \bibinfo{journal}{{\em Perspectives on psychological science\/}}
  \bibinfo{volume}{6}, \bibinfo{number}{1} (\bibinfo{year}{2011}),
  \bibinfo{pages}{3--5}.
\newblock


\bibitem[\protect\citeauthoryear{Cochran, D'Antoni, Livshits, Molnar, and
  Veanes}{Cochran et~al\mbox{.}}{2015a}]%
        {Cochran:2015:PBP:2676726.2676973}
\bibfield{author}{\bibinfo{person}{Robert~A. Cochran}, \bibinfo{person}{Loris
  D'Antoni}, \bibinfo{person}{Benjamin Livshits}, \bibinfo{person}{David
  Molnar}, {and} \bibinfo{person}{Margus Veanes}.}
  \bibinfo{year}{2015}\natexlab{a}.
\newblock \showarticletitle{Program Boosting: Program Synthesis via
  Crowd-Sourcing}. In \bibinfo{booktitle}{{\em Proceedings of the 42Nd Annual
  ACM SIGPLAN-SIGACT Symposium on Principles of Programming Languages}} {\em
  (\bibinfo{series}{POPL '15})}. \bibinfo{publisher}{ACM},
  \bibinfo{address}{New York, NY, USA}, \bibinfo{pages}{677--688}.
\newblock
\showISBNx{978-1-4503-3300-9}
\showDOI{%
\url{http://dx.doi.org/10.1145/2676726.2676973}}


\bibitem[\protect\citeauthoryear{Cochran, D'Antoni, Livshits, Molnar, and
  Veanes}{Cochran et~al\mbox{.}}{2015b}]%
        {Cochran:2015:PBP:2775051.2676973}
\bibfield{author}{\bibinfo{person}{Robert~A. Cochran}, \bibinfo{person}{Loris
  D'Antoni}, \bibinfo{person}{Benjamin Livshits}, \bibinfo{person}{David
  Molnar}, {and} \bibinfo{person}{Margus Veanes}.}
  \bibinfo{year}{2015}\natexlab{b}.
\newblock \showarticletitle{Program Boosting: Program Synthesis via
  Crowd-Sourcing}.
\newblock \bibinfo{journal}{{\em SIGPLAN Not.\/}} \bibinfo{volume}{50},
  \bibinfo{number}{1} (\bibinfo{date}{Jan.} \bibinfo{year}{2015}),
  \bibinfo{pages}{677--688}.
\newblock
\showISSN{0362-1340}
\showDOI{%
\url{http://dx.doi.org/10.1145/2775051.2676973}}


\bibitem[\protect\citeauthoryear{Cohen}{Cohen}{1989}]%
        {cohen1989developing}
\bibfield{author}{\bibinfo{person}{Bernard~P Cohen}.}
  \bibinfo{year}{1989}\natexlab{}.
\newblock \bibinfo{booktitle}{{\em Developing sociological knowledge: Theory
  and method}}.
\newblock \bibinfo{publisher}{Wadsworth Pub Co}.
\newblock


\bibitem[\protect\citeauthoryear{Cooper, Hedges, and Valentine}{Cooper
  et~al\mbox{.}}{2009}]%
        {cooper2009handbook}
\bibfield{author}{\bibinfo{person}{Harris Cooper}, \bibinfo{person}{Larry~V
  Hedges}, {and} \bibinfo{person}{Jeffrey~C Valentine}.}
  \bibinfo{year}{2009}\natexlab{}.
\newblock \bibinfo{booktitle}{{\em The handbook of research synthesis and
  meta-analysis}}.
\newblock \bibinfo{publisher}{Russell Sage Foundation}.
\newblock


\bibitem[\protect\citeauthoryear{Dabbish, Stuart, Tsay, and Herbsleb}{Dabbish
  et~al\mbox{.}}{2012}]%
        {dabbish2012social}
\bibfield{author}{\bibinfo{person}{Laura Dabbish}, \bibinfo{person}{Colleen
  Stuart}, \bibinfo{person}{Jason Tsay}, {and} \bibinfo{person}{Jim Herbsleb}.}
  \bibinfo{year}{2012}\natexlab{}.
\newblock \showarticletitle{Social coding in GitHub: transparency and
  collaboration in an open software repository}. In \bibinfo{booktitle}{{\em
  Proceedings of the ACM 2012 conference on Computer Supported Cooperative
  Work}}. ACM, \bibinfo{pages}{1277--1286}.
\newblock


\bibitem[\protect\citeauthoryear{Dolstra, Vliegendhart, and Pouwelse}{Dolstra
  et~al\mbox{.}}{2013}]%
        {6569745}
\bibfield{author}{\bibinfo{person}{E. Dolstra}, \bibinfo{person}{R.
  Vliegendhart}, {and} \bibinfo{person}{J. Pouwelse}.}
  \bibinfo{year}{2013}\natexlab{}.
\newblock \showarticletitle{Crowdsourcing GUI Tests}. In
  \bibinfo{booktitle}{{\em Software Testing, Verification and Validation
  (ICST), 2013 IEEE Sixth International Conference on}}.
  \bibinfo{pages}{332--341}.
\newblock
\showDOI{%
\url{http://dx.doi.org/10.1109/ICST.2013.44}}


\bibitem[\protect\citeauthoryear{Easterbrook, Singer, Storey, and
  Damian}{Easterbrook et~al\mbox{.}}{2008}]%
        {easterbrook2008selecting}
\bibfield{author}{\bibinfo{person}{Steve Easterbrook}, \bibinfo{person}{Janice
  Singer}, \bibinfo{person}{Margaret-Anne Storey}, {and}
  \bibinfo{person}{Daniela Damian}.} \bibinfo{year}{2008}\natexlab{}.
\newblock \showarticletitle{Selecting empirical methods for software
  engineering research}.
\newblock In \bibinfo{booktitle}{{\em Guide to advanced empirical software
  engineering}}. \bibinfo{publisher}{Springer}, \bibinfo{pages}{285--311}.
\newblock


\bibitem[\protect\citeauthoryear{Goel, Nikzad, and Singla}{Goel
  et~al\mbox{.}}{2014}]%
        {goel2014mechanism}
\bibfield{author}{\bibinfo{person}{Gagan Goel}, \bibinfo{person}{Afshin
  Nikzad}, {and} \bibinfo{person}{Adish Singla}.}
  \bibinfo{year}{2014}\natexlab{}.
\newblock \showarticletitle{Mechanism design for crowdsourcing markets with
  heterogeneous tasks}. In \bibinfo{booktitle}{{\em Second AAAI Conference on
  Human Computation and Crowdsourcing}}.
\newblock


\bibitem[\protect\citeauthoryear{Gousios}{Gousios}{2013}]%
        {Gousi13GHTorrent}
\bibfield{author}{\bibinfo{person}{Georgios Gousios}.}
  \bibinfo{year}{2013}\natexlab{}.
\newblock \showarticletitle{The GHTorrent dataset and tool suite}. In
  \bibinfo{booktitle}{{\em Proceedings of the 10th Working Conference on Mining
  Software Repositories}} {\em (\bibinfo{series}{MSR '13})}.
  \bibinfo{publisher}{IEEE Press}, \bibinfo{address}{Piscataway, NJ, USA},
  \bibinfo{pages}{233--236}.
\newblock
\showISBNx{978-1-4673-2936-1}
\showURL{%
\url{http://dl.acm.org/citation.cfm?id=2487085.2487132}}


\bibitem[\protect\citeauthoryear{Gousios, Pinzger, and Deursen}{Gousios
  et~al\mbox{.}}{2014}]%
        {gousios2014exploratory}
\bibfield{author}{\bibinfo{person}{Georgios Gousios}, \bibinfo{person}{Martin
  Pinzger}, {and} \bibinfo{person}{Arie~van Deursen}.}
  \bibinfo{year}{2014}\natexlab{}.
\newblock \showarticletitle{An exploratory study of the pull-based software
  development model}. In \bibinfo{booktitle}{{\em Proceedings of the 36th
  International Conference on Software Engineering}}. ACM,
  \bibinfo{pages}{345--355}.
\newblock


\bibitem[\protect\citeauthoryear{Gousios, Pinzger, and van Deursen}{Gousios
  et~al\mbox{.}}{2013}]%
        {gousios2013exploration}
\bibfield{author}{\bibinfo{person}{G Gousios}, \bibinfo{person}{M Pinzger},
  {and} \bibinfo{person}{A van Deursen}.} \bibinfo{year}{2013}\natexlab{}.
\newblock \showarticletitle{An exploration of the pull-based software
  development model}. ICSE.
\newblock


\bibitem[\protect\citeauthoryear{Gousios, Storey, and Bacchelli}{Gousios
  et~al\mbox{.}}{2016}]%
        {gousios2016work}
\bibfield{author}{\bibinfo{person}{Georgios Gousios},
  \bibinfo{person}{Margaret-Anne Storey}, {and} \bibinfo{person}{Alberto
  Bacchelli}.} \bibinfo{year}{2016}\natexlab{}.
\newblock \showarticletitle{Work practices and challenges in pull-based
  development: the contributor's perspective}. In \bibinfo{booktitle}{{\em
  Proceedings of the 38th International Conference on Software Engineering}}.
  ACM, \bibinfo{pages}{285--296}.
\newblock


\bibitem[\protect\citeauthoryear{Gousios and Zaidman}{Gousios and
  Zaidman}{2014}]%
        {gousios2014dataset}
\bibfield{author}{\bibinfo{person}{Georgios Gousios} {and}
  \bibinfo{person}{Andy Zaidman}.} \bibinfo{year}{2014}\natexlab{}.
\newblock \showarticletitle{A dataset for pull-based development research}. In
  \bibinfo{booktitle}{{\em Proceedings of the 11th Working Conference on Mining
  Software Repositories}}. ACM, \bibinfo{pages}{368--371}.
\newblock


\bibitem[\protect\citeauthoryear{Gousios, Zaidman, Storey, and
  Van~Deursen}{Gousios et~al\mbox{.}}{2015}]%
        {gousios2015work}
\bibfield{author}{\bibinfo{person}{Georgios Gousios}, \bibinfo{person}{Andy
  Zaidman}, \bibinfo{person}{Margaret-Anne Storey}, {and} \bibinfo{person}{Arie
  Van~Deursen}.} \bibinfo{year}{2015}\natexlab{}.
\newblock \showarticletitle{Work practices and challenges in pull-based
  development: the integrator's perspective}. In \bibinfo{booktitle}{{\em
  Proceedings of the 37th International Conference on Software
  Engineering-Volume 1}}. IEEE Press, \bibinfo{pages}{358--368}.
\newblock


\bibitem[\protect\citeauthoryear{Guba, Lincoln, et~al\mbox{.}}{Guba
  et~al\mbox{.}}{1994}]%
        {guba1994competing}
\bibfield{author}{\bibinfo{person}{Egon~G Guba}, \bibinfo{person}{Yvonna~S
  Lincoln}, {and} \bibinfo{person}{others}.} \bibinfo{year}{1994}\natexlab{}.
\newblock \showarticletitle{Competing paradigms in qualitative research}.
\newblock \bibinfo{journal}{{\em Handbook of qualitative research\/}}
  \bibinfo{volume}{2}, \bibinfo{number}{163-194} (\bibinfo{year}{1994}),
  \bibinfo{pages}{105}.
\newblock


\bibitem[\protect\citeauthoryear{Guzzi, Bacchelli, Lanza, Pinzger, and
  Van~Deursen}{Guzzi et~al\mbox{.}}{2013}]%
        {guzzi2013communication}
\bibfield{author}{\bibinfo{person}{Anja Guzzi}, \bibinfo{person}{Alberto
  Bacchelli}, \bibinfo{person}{Michele Lanza}, \bibinfo{person}{Martin
  Pinzger}, {and} \bibinfo{person}{Arie Van~Deursen}.}
  \bibinfo{year}{2013}\natexlab{}.
\newblock \showarticletitle{Communication in open source software development
  mailing lists}. In \bibinfo{booktitle}{{\em Mining Software Repositories
  (MSR), 2013 10th IEEE Working Conference on}}. IEEE,
  \bibinfo{pages}{277--286}.
\newblock


\bibitem[\protect\citeauthoryear{Hall}{Hall}{1999}]%
        {hall1999correlation}
\bibfield{author}{\bibinfo{person}{Mark~A Hall}.}
  \bibinfo{year}{1999}\natexlab{}.
\newblock {\em \bibinfo{title}{Correlation-based feature selection for machine
  learning}}.
\newblock \bibinfo{thesistype}{Ph.D. Dissertation}. \bibinfo{school}{The
  University of Waikato}.
\newblock


\bibitem[\protect\citeauthoryear{Heikinheimo and Ukkonen}{Heikinheimo and
  Ukkonen}{2013}]%
        {Heikinheimo2013}
\bibfield{author}{\bibinfo{person}{Hannes Heikinheimo} {and} \bibinfo{person}{a
  Ukkonen}.} \bibinfo{year}{2013}\natexlab{}.
\newblock \showarticletitle{{The Crowd-Median Algorithm}}.
\newblock \bibinfo{journal}{{\em First AAAI Conference on Human {\ldots}\/}}
  (\bibinfo{year}{2013}), \bibinfo{pages}{69--77}.
\newblock
\showISBNx{978-1-57735-607-3}
\showURL{%
\url{http://www.aaai.org/ocs/index.php/HCOMP/HCOMP13/paper/view/7513}}


\bibitem[\protect\citeauthoryear{Ipeirotis}{Ipeirotis}{2010}]%
        {ipeirotis2010demographics}
\bibfield{author}{\bibinfo{person}{Panagiotis~G Ipeirotis}.}
  \bibinfo{year}{2010}\natexlab{}.
\newblock \showarticletitle{Demographics of mechanical turk}.
\newblock  (\bibinfo{year}{2010}).
\newblock


\bibitem[\protect\citeauthoryear{Kalliamvakou, Gousios, Blincoe, Singer,
  German, and Damian}{Kalliamvakou et~al\mbox{.}}{2014}]%
        {kalliamvakou2014promises}
\bibfield{author}{\bibinfo{person}{Eirini Kalliamvakou},
  \bibinfo{person}{Georgios Gousios}, \bibinfo{person}{Kelly Blincoe},
  \bibinfo{person}{Leif Singer}, \bibinfo{person}{Daniel~M German}, {and}
  \bibinfo{person}{Daniela Damian}.} \bibinfo{year}{2014}\natexlab{}.
\newblock \showarticletitle{The promises and perils of mining GitHub}. In
  \bibinfo{booktitle}{{\em Proceedings of the 11th working conference on mining
  software repositories}}. ACM, \bibinfo{pages}{92--101}.
\newblock


\bibitem[\protect\citeauthoryear{Kim, Yun, and Yi}{Kim et~al\mbox{.}}{2012}]%
        {Kim:2012:FOR:2442576.2442591}
\bibfield{author}{\bibinfo{person}{Sung-Hee Kim}, \bibinfo{person}{Hyokun Yun},
  {and} \bibinfo{person}{Ji~Soo Yi}.} \bibinfo{year}{2012}\natexlab{}.
\newblock \showarticletitle{How to Filter out Random Clickers in a
  Crowdsourcing-based Study?}. In \bibinfo{booktitle}{{\em Proceedings of the
  2012 BELIV Workshop: Beyond Time and Errors - Novel Evaluation Methods for
  Visualization}} {\em (\bibinfo{series}{BELIV '12})}.
  \bibinfo{publisher}{ACM}, \bibinfo{address}{New York, NY, USA}, Article
  \bibinfo{articleno}{15}, \bibinfo{numpages}{7}~pages.
\newblock
\showISBNx{978-1-4503-1791-7}
\showDOI{%
\url{http://dx.doi.org/10.1145/2442576.2442591}}


\bibitem[\protect\citeauthoryear{Kittur, Chi, and Suh}{Kittur
  et~al\mbox{.}}{2008}]%
        {kittur2008crowdsourcing}
\bibfield{author}{\bibinfo{person}{Aniket Kittur}, \bibinfo{person}{Ed~H Chi},
  {and} \bibinfo{person}{Bongwon Suh}.} \bibinfo{year}{2008}\natexlab{}.
\newblock \showarticletitle{Crowdsourcing user studies with Mechanical Turk}.
  In \bibinfo{booktitle}{{\em Proceedings of the SIGCHI conference on human
  factors in computing systems}}. ACM, \bibinfo{pages}{453--456}.
\newblock


\bibitem[\protect\citeauthoryear{Krosnick}{Krosnick}{1991}]%
        {krosnick1991response}
\bibfield{author}{\bibinfo{person}{Jon~A Krosnick}.}
  \bibinfo{year}{1991}\natexlab{}.
\newblock \showarticletitle{Response strategies for coping with the cognitive
  demands of attitude measures in surveys}.
\newblock \bibinfo{journal}{{\em Applied cognitive psychology\/}}
  \bibinfo{volume}{5}, \bibinfo{number}{3} (\bibinfo{year}{1991}),
  \bibinfo{pages}{213--236}.
\newblock


\bibitem[\protect\citeauthoryear{L{\'a}zaro and Marcos}{L{\'a}zaro and
  Marcos}{2006}]%
        {lazaro2006approach}
\bibfield{author}{\bibinfo{person}{Mar{\'\i}a L{\'a}zaro} {and}
  \bibinfo{person}{Esperanza Marcos}.} \bibinfo{year}{2006}\natexlab{}.
\newblock \showarticletitle{An Approach to the Integration of Qualitative and
  Quantitative Research Methods in Software Engineering Research.}
\newblock \bibinfo{journal}{{\em PhiSE\/}}  \bibinfo{volume}{240}
  (\bibinfo{year}{2006}).
\newblock


\bibitem[\protect\citeauthoryear{Little}{Little}{2009}]%
        {5295247}
\bibfield{author}{\bibinfo{person}{G. Little}.}
  \bibinfo{year}{2009}\natexlab{}.
\newblock \showarticletitle{TurKit: Tools for iterative tasks on mechanical
  turk}. In \bibinfo{booktitle}{{\em Visual Languages and Human-Centric
  Computing, 2009. VL/HCC 2009. IEEE Symposium on}}. \bibinfo{pages}{252--253}.
\newblock
\showISSN{1943-6092}
\showDOI{%
\url{http://dx.doi.org/10.1109/VLHCC.2009.5295247}}


\bibitem[\protect\citeauthoryear{Mao, Kamar, Chen, Horvitz, Schwamb, Lintott,
  and Smith}{Mao et~al\mbox{.}}{2013a}]%
        {mao2013volunteering}
\bibfield{author}{\bibinfo{person}{Andrew Mao}, \bibinfo{person}{Ece Kamar},
  \bibinfo{person}{Yiling Chen}, \bibinfo{person}{Eric Horvitz},
  \bibinfo{person}{Megan~E Schwamb}, \bibinfo{person}{Chris~J Lintott}, {and}
  \bibinfo{person}{Arfon~M Smith}.} \bibinfo{year}{2013}\natexlab{a}.
\newblock \showarticletitle{Volunteering vs. work for pay: incentives and
  tradeoffs in crowdsourcing}. In \bibinfo{booktitle}{{\em Conf. on Human
  Computation}}.
\newblock


\bibitem[\protect\citeauthoryear{Mao, Capra, Harman, and Jia}{Mao
  et~al\mbox{.}}{2015}]%
        {mao2015survey}
\bibfield{author}{\bibinfo{person}{Ke Mao}, \bibinfo{person}{Licia Capra},
  \bibinfo{person}{Mark Harman}, {and} \bibinfo{person}{Yue Jia}.}
  \bibinfo{year}{2015}\natexlab{}.
\newblock \showarticletitle{A survey of the use of crowdsourcing in software
  engineering}.
\newblock \bibinfo{journal}{{\em RN\/}} \bibinfo{volume}{15},
  \bibinfo{number}{01} (\bibinfo{year}{2015}).
\newblock


\bibitem[\protect\citeauthoryear{Mao, Yang, Li, and Harman}{Mao
  et~al\mbox{.}}{2013b}]%
        {Mao:2013:PCS:2486788.2486963}
\bibfield{author}{\bibinfo{person}{Ke Mao}, \bibinfo{person}{Ye Yang},
  \bibinfo{person}{Mingshu Li}, {and} \bibinfo{person}{Mark Harman}.}
  \bibinfo{year}{2013}\natexlab{b}.
\newblock \showarticletitle{Pricing Crowdsourcing-based Software Development
  Tasks}. In \bibinfo{booktitle}{{\em Proceedings of the 2013 International
  Conference on Software Engineering}} {\em (\bibinfo{series}{ICSE '13})}.
  \bibinfo{publisher}{IEEE Press}, \bibinfo{address}{Piscataway, NJ, USA},
  \bibinfo{pages}{1205--1208}.
\newblock
\showISBNx{978-1-4673-3076-3}
\showURL{%
\url{http://dl.acm.org/citation.cfm?id=2486788.2486963}}


\bibitem[\protect\citeauthoryear{Marlow, Dabbish, and Herbsleb}{Marlow
  et~al\mbox{.}}{2013}]%
        {marlow2013impression}
\bibfield{author}{\bibinfo{person}{Jennifer Marlow}, \bibinfo{person}{Laura
  Dabbish}, {and} \bibinfo{person}{Jim Herbsleb}.}
  \bibinfo{year}{2013}\natexlab{}.
\newblock \showarticletitle{Impression formation in online peer production:
  activity traces and personal profiles in github}. In \bibinfo{booktitle}{{\em
  Proceedings of the 2013 conference on Computer supported cooperative work}}.
  ACM, \bibinfo{pages}{117--128}.
\newblock


\bibitem[\protect\citeauthoryear{Mason}{Mason}{2013}]%
        {Mason2013}
\bibfield{author}{\bibinfo{person}{Winter Mason}.}
  \bibinfo{year}{2013}\natexlab{}.
\newblock \bibinfo{booktitle}{{\em Handbook of Human Computation}}.
\newblock \bibinfo{publisher}{Springer New York}, \bibinfo{address}{New York,
  NY}, Chapter Collective Search as Human Computation,
  \bibinfo{pages}{463--474}.
\newblock
\showISBNx{978-1-4614-8806-4}
\showDOI{%
\url{http://dx.doi.org/10.1007/978-1-4614-8806-4_35}}


\bibitem[\protect\citeauthoryear{Mason and Watts}{Mason and Watts}{2010}]%
        {mason2010financial}
\bibfield{author}{\bibinfo{person}{Winter Mason} {and}
  \bibinfo{person}{Duncan~J Watts}.} \bibinfo{year}{2010}\natexlab{}.
\newblock \showarticletitle{Financial incentives and the performance of
  crowds}.
\newblock \bibinfo{journal}{{\em ACM SigKDD Explorations Newsletter\/}}
  \bibinfo{volume}{11}, \bibinfo{number}{2} (\bibinfo{year}{2010}),
  \bibinfo{pages}{100--108}.
\newblock


\bibitem[\protect\citeauthoryear{Mason and Suri}{Mason and Suri}{2011}]%
        {mason2011use}
\bibfield{author}{\bibinfo{person}{Winter~A Mason} {and}
  \bibinfo{person}{Siddharth Suri}.} \bibinfo{year}{2011}\natexlab{}.
\newblock \showarticletitle{How to use mechanical turk for cognitive science
  research}. In \bibinfo{booktitle}{{\em Proceedings of the 33rd annual
  conference of the cognitive science society}}. \bibinfo{pages}{66--67}.
\newblock


\bibitem[\protect\citeauthoryear{Mathew, Agrawal, and Menzies}{Mathew
  et~al\mbox{.}}{}]%
        {mathew17}
\bibfield{author}{\bibinfo{person}{George Mathew}, \bibinfo{person}{Amritanshu
  Agrawal}, {and} \bibinfo{person}{Tim Menzies}.}
\newblock \bibinfo{title}{Trends in Topics in Software Engineering Conferences,
  1992 to 2016}.
\newblock \bibinfo{howpublished}{Submitted to IST. Available from
  \url{http://tiny.cc/citemap}}.   (\bibinfo{year}{????}).
\newblock


\bibitem[\protect\citeauthoryear{McDonald and Goggins}{McDonald and
  Goggins}{2013}]%
        {mcdonald2013performance}
\bibfield{author}{\bibinfo{person}{Nora McDonald} {and} \bibinfo{person}{Sean
  Goggins}.} \bibinfo{year}{2013}\natexlab{}.
\newblock \showarticletitle{Performance and participation in open source
  software on github}. In \bibinfo{booktitle}{{\em CHI'13 Extended Abstracts on
  Human Factors in Computing Systems}}. ACM, \bibinfo{pages}{139--144}.
\newblock


\bibitem[\protect\citeauthoryear{Menzies, Kocag{\"{u}}neli, Minku, Peters, and
  Turhan}{Menzies et~al\mbox{.}}{2015}]%
        {minku15}
\bibfield{author}{\bibinfo{person}{Tim Menzies}, \bibinfo{person}{Ekrem
  Kocag{\"{u}}neli}, \bibinfo{person}{Leandro Minku}, \bibinfo{person}{Fayola
  Peters}, {and} \bibinfo{person}{Burak Turhan}.}
  \bibinfo{year}{2015}\natexlab{}.
\newblock \showarticletitle{{Chapter 20 - Ensembles of Learning Machines}}.
\newblock In \bibinfo{booktitle}{{\em Sharing Data and Models in Software
  Engineering}}. \bibinfo{pages}{239--265}.
\newblock
\showISBNx{978-0-12-417295-1}
\showDOI{%
\url{http://dx.doi.org/10.1016/B978-0-12-417295-1.00020-5}}


\bibitem[\protect\citeauthoryear{Nebeling, Speicher, and Norrie}{Nebeling
  et~al\mbox{.}}{2013}]%
        {Nebeling:2013:CGT:2494603.2480303}
\bibfield{author}{\bibinfo{person}{Michael Nebeling},
  \bibinfo{person}{Maximilian Speicher}, {and} \bibinfo{person}{Moira~C.
  Norrie}.} \bibinfo{year}{2013}\natexlab{}.
\newblock \showarticletitle{CrowdStudy: General Toolkit for Crowdsourced
  Evaluation of Web Interfaces}. In \bibinfo{booktitle}{{\em Proceedings of the
  5th ACM SIGCHI Symposium on Engineering Interactive Computing Systems}} {\em
  (\bibinfo{series}{EICS '13})}. \bibinfo{publisher}{ACM},
  \bibinfo{address}{New York, NY, USA}, \bibinfo{pages}{255--264}.
\newblock
\showISBNx{978-1-4503-2138-9}
\showDOI{%
\url{http://dx.doi.org/10.1145/2494603.2480303}}


\bibitem[\protect\citeauthoryear{O'Neil}{O'Neil}{2016}]%
        {o2016weapons}
\bibfield{author}{\bibinfo{person}{Cathy O'Neil}.}
  \bibinfo{year}{2016}\natexlab{}.
\newblock \bibinfo{booktitle}{{\em Weapons of Math Destruction: How Big Data
  Increases Inequality and Threatens Democracy}}.
\newblock \bibinfo{publisher}{Crown Publishing Group (NY)}.
\newblock


\bibitem[\protect\citeauthoryear{Padhye, Mani, and Sinha}{Padhye
  et~al\mbox{.}}{2014}]%
        {padhye2014study}
\bibfield{author}{\bibinfo{person}{Rohan Padhye}, \bibinfo{person}{Senthil
  Mani}, {and} \bibinfo{person}{Vibha~Singhal Sinha}.}
  \bibinfo{year}{2014}\natexlab{}.
\newblock \showarticletitle{A study of external community contribution to
  open-source projects on GitHub}. In \bibinfo{booktitle}{{\em Proceedings of
  the 11th Working Conference on Mining Software Repositories}}. ACM,
  \bibinfo{pages}{332--335}.
\newblock


\bibitem[\protect\citeauthoryear{Paolacci, Chandler, and Ipeirotis}{Paolacci
  et~al\mbox{.}}{2010}]%
        {paolacci2010running}
\bibfield{author}{\bibinfo{person}{Gabriele Paolacci}, \bibinfo{person}{Jesse
  Chandler}, {and} \bibinfo{person}{Panagiotis~G Ipeirotis}.}
  \bibinfo{year}{2010}\natexlab{}.
\newblock \showarticletitle{Running experiments on amazon mechanical turk}.
\newblock \bibinfo{journal}{{\em Judgment and Decision making\/}}
  \bibinfo{volume}{5}, \bibinfo{number}{5} (\bibinfo{year}{2010}),
  \bibinfo{pages}{411--419}.
\newblock


\bibitem[\protect\citeauthoryear{Pham, Singer, Liskin, Figueira~Filho, and
  Schneider}{Pham et~al\mbox{.}}{2013}]%
        {pham2013creating}
\bibfield{author}{\bibinfo{person}{Raphael Pham}, \bibinfo{person}{Leif
  Singer}, \bibinfo{person}{Olga Liskin}, \bibinfo{person}{Fernando
  Figueira~Filho}, {and} \bibinfo{person}{Kurt Schneider}.}
  \bibinfo{year}{2013}\natexlab{}.
\newblock \showarticletitle{Creating a shared understanding of testing culture
  on a social coding site}. In \bibinfo{booktitle}{{\em 2013 35th International
  Conference on Software Engineering (ICSE)}}. IEEE, \bibinfo{pages}{112--121}.
\newblock


\bibitem[\protect\citeauthoryear{Pletea, Vasilescu, and Serebrenik}{Pletea
  et~al\mbox{.}}{2014}]%
        {pletea2014security}
\bibfield{author}{\bibinfo{person}{Daniel Pletea}, \bibinfo{person}{Bogdan
  Vasilescu}, {and} \bibinfo{person}{Alexander Serebrenik}.}
  \bibinfo{year}{2014}\natexlab{}.
\newblock \showarticletitle{Security and emotion: sentiment analysis of
  security discussions on GitHub}. In \bibinfo{booktitle}{{\em Proceedings of
  the 11th working conference on mining software repositories}}. ACM,
  \bibinfo{pages}{348--351}.
\newblock


\bibitem[\protect\citeauthoryear{Rahman and Roy}{Rahman and Roy}{2014}]%
        {rahman2014insight}
\bibfield{author}{\bibinfo{person}{Mohammad~Masudur Rahman} {and}
  \bibinfo{person}{Chanchal~K Roy}.} \bibinfo{year}{2014}\natexlab{}.
\newblock \showarticletitle{An insight into the pull requests of github}. In
  \bibinfo{booktitle}{{\em Proceedings of the 11th Working Conference on Mining
  Software Repositories}}. ACM, \bibinfo{pages}{364--367}.
\newblock


\bibitem[\protect\citeauthoryear{Ray, Posnett, Filkov, and Devanbu}{Ray
  et~al\mbox{.}}{2014}]%
        {ray2014large}
\bibfield{author}{\bibinfo{person}{Baishakhi Ray}, \bibinfo{person}{Daryl
  Posnett}, \bibinfo{person}{Vladimir Filkov}, {and} \bibinfo{person}{Premkumar
  Devanbu}.} \bibinfo{year}{2014}\natexlab{}.
\newblock \showarticletitle{A large scale study of programming languages and
  code quality in github}. In \bibinfo{booktitle}{{\em Proceedings of the 22nd
  ACM SIGSOFT International Symposium on Foundations of Software Engineering}}.
  ACM, \bibinfo{pages}{155--165}.
\newblock


\bibitem[\protect\citeauthoryear{Sadowski, Stolee, and Elbaum}{Sadowski
  et~al\mbox{.}}{2015}]%
        {stoleeFSE2015}
\bibfield{author}{\bibinfo{person}{Caitlin Sadowski},
  \bibinfo{person}{Kathryn~T. Stolee}, {and} \bibinfo{person}{Sebastian
  Elbaum}.} \bibinfo{year}{2015}\natexlab{}.
\newblock \showarticletitle{How Developers Search for Code: A Case Study}. In
  \bibinfo{booktitle}{{\em European Software Engineering Conference and the ACM
  SIGSOFT Symposium on the Foundations of Software Engineering (ESEC/FSE)}}.
\newblock


\bibitem[\protect\citeauthoryear{Sale, Lohfeld, and Brazil}{Sale
  et~al\mbox{.}}{2002}]%
        {sale2002revisiting}
\bibfield{author}{\bibinfo{person}{Joanna~EM Sale}, \bibinfo{person}{Lynne~H
  Lohfeld}, {and} \bibinfo{person}{Kevin Brazil}.}
  \bibinfo{year}{2002}\natexlab{}.
\newblock \showarticletitle{Revisiting the quantitative-qualitative debate:
  Implications for mixed-methods research}.
\newblock \bibinfo{journal}{{\em Quality and quantity\/}} \bibinfo{volume}{36},
  \bibinfo{number}{1} (\bibinfo{year}{2002}), \bibinfo{pages}{43--53}.
\newblock


\bibitem[\protect\citeauthoryear{Schiller and Ernst}{Schiller and
  Ernst}{2012}]%
        {Schiller:2012:RBW:2398857.2384624}
\bibfield{author}{\bibinfo{person}{Todd~W. Schiller} {and}
  \bibinfo{person}{Michael~D. Ernst}.} \bibinfo{year}{2012}\natexlab{}.
\newblock \showarticletitle{Reducing the Barriers to Writing Verified
  Specifications}.
\newblock \bibinfo{journal}{{\em SIGPLAN Not.\/}} \bibinfo{volume}{47},
  \bibinfo{number}{10} (\bibinfo{date}{Oct.} \bibinfo{year}{2012}),
  \bibinfo{pages}{95--112}.
\newblock
\showISSN{0362-1340}
\showDOI{%
\url{http://dx.doi.org/10.1145/2398857.2384624}}


\bibitem[\protect\citeauthoryear{Seaman}{Seaman}{1999}]%
        {seaman1999qualitative}
\bibfield{author}{\bibinfo{person}{Carolyn~B. Seaman}.}
  \bibinfo{year}{1999}\natexlab{}.
\newblock \showarticletitle{Qualitative methods in empirical studies of
  software engineering}.
\newblock \bibinfo{journal}{{\em IEEE Transactions on software engineering\/}}
  \bibinfo{volume}{25}, \bibinfo{number}{4} (\bibinfo{year}{1999}),
  \bibinfo{pages}{557--572}.
\newblock


\bibitem[\protect\citeauthoryear{Shiel}{Shiel}{2013}]%
        {shiel2013conflict}
\bibfield{author}{\bibinfo{person}{Annie Shiel}.}
  \bibinfo{year}{2013}\natexlab{}.
\newblock \showarticletitle{Conflict Crowdsourcing: Harnessing the power of
  crowdsourcing for organizations working in conflict}.
\newblock  (\bibinfo{year}{2013}).
\newblock


\bibitem[\protect\citeauthoryear{Shull, Basili, Carver, Maldonado, Travassos,
  Mendon{\c{c}}a, and Fabbri}{Shull et~al\mbox{.}}{2002}]%
        {shull02}
\bibfield{author}{\bibinfo{person}{Forrest Shull}, \bibinfo{person}{Victor
  Basili}, \bibinfo{person}{Jeffrey Carver}, \bibinfo{person}{Jos{\'{e}}~C
  Maldonado}, \bibinfo{person}{Guilherme~Horta Travassos},
  \bibinfo{person}{Manoel Mendon{\c{c}}a}, {and} \bibinfo{person}{Sandra
  Fabbri}.} \bibinfo{year}{2002}\natexlab{}.
\newblock \showarticletitle{{Replicating Software Engineering Experiments:
  Addressing the Tacit Knowledge Problem}}. In \bibinfo{booktitle}{{\em ISESE
  '02: Proceedings of the 2002 International Symposium on Empirical Software
  Engineering}}. \bibinfo{publisher}{IEEE Computer Society},
  \bibinfo{address}{Washington, DC, USA}, \bibinfo{pages}{7}.
\newblock
\showISBNx{0-7695-1796-X}


\bibitem[\protect\citeauthoryear{Shull, Carver, Vegas, and Juristo}{Shull
  et~al\mbox{.}}{2008}]%
        {Shull:2008:RRE:1361580.1361587}
\bibfield{author}{\bibinfo{person}{Forrest~J. Shull},
  \bibinfo{person}{Jeffrey~C. Carver}, \bibinfo{person}{Sira Vegas}, {and}
  \bibinfo{person}{Natalia Juristo}.} \bibinfo{year}{2008}\natexlab{}.
\newblock \showarticletitle{The Role of Replications in Empirical Software
  Engineering}.
\newblock \bibinfo{journal}{{\em Empirical Softw. Engg.\/}}
  \bibinfo{volume}{13}, \bibinfo{number}{2} (\bibinfo{date}{April}
  \bibinfo{year}{2008}), \bibinfo{pages}{211--218}.
\newblock
\showISSN{1382-3256}
\showDOI{%
\url{http://dx.doi.org/10.1007/s10664-008-9060-1}}


\bibitem[\protect\citeauthoryear{Siegmund, Siegmund, and Apel}{Siegmund
  et~al\mbox{.}}{2015}]%
        {siegmund2015views}
\bibfield{author}{\bibinfo{person}{Janet Siegmund}, \bibinfo{person}{Norbert
  Siegmund}, {and} \bibinfo{person}{Sven Apel}.}
  \bibinfo{year}{2015}\natexlab{}.
\newblock \showarticletitle{Views on internal and external validity in
  empirical software engineering}. In \bibinfo{booktitle}{{\em Software
  Engineering (ICSE), 2015 IEEE/ACM 37th IEEE International Conference on}},
  Vol.~\bibinfo{volume}{1}. IEEE, \bibinfo{pages}{9--19}.
\newblock


\bibitem[\protect\citeauthoryear{Sjoberg, Dyba, and Jorgensen}{Sjoberg
  et~al\mbox{.}}{2007}]%
        {sjoberg2007future}
\bibfield{author}{\bibinfo{person}{Dag~IK Sjoberg}, \bibinfo{person}{Tore
  Dyba}, {and} \bibinfo{person}{Magne Jorgensen}.}
  \bibinfo{year}{2007}\natexlab{}.
\newblock \showarticletitle{The future of empirical methods in software
  engineering research}. In \bibinfo{booktitle}{{\em Future of Software
  Engineering, 2007. FOSE'07}}. IEEE, \bibinfo{pages}{358--378}.
\newblock


\bibitem[\protect\citeauthoryear{Stolee, Saylor, and Lund}{Stolee
  et~al\mbox{.}}{2015}]%
        {stolee2015exploring}
\bibfield{author}{\bibinfo{person}{Kathryn~T Stolee}, \bibinfo{person}{James
  Saylor}, {and} \bibinfo{person}{Trevor Lund}.}
  \bibinfo{year}{2015}\natexlab{}.
\newblock \showarticletitle{Exploring the benefits of using redundant responses
  in crowdsourced evaluations}. In \bibinfo{booktitle}{{\em Proceedings of the
  Second International Workshop on CrowdSourcing in Software Engineering}}.
  IEEE Press, \bibinfo{pages}{38--44}.
\newblock


\bibitem[\protect\citeauthoryear{Takhteyev and Hilts}{Takhteyev and
  Hilts}{2010}]%
        {takhteyev2010investigating}
\bibfield{author}{\bibinfo{person}{Yuri Takhteyev} {and}
  \bibinfo{person}{Andrew Hilts}.} \bibinfo{year}{2010}\natexlab{}.
\newblock \bibinfo{title}{Investigating the geography of open source software
  through GitHub}.
\newblock   (\bibinfo{year}{2010}).
\newblock


\bibitem[\protect\citeauthoryear{Thung, Bissyande, Lo, and Jiang}{Thung
  et~al\mbox{.}}{2013}]%
        {thung2013network}
\bibfield{author}{\bibinfo{person}{Ferdian Thung}, \bibinfo{person}{Tegawende~F
  Bissyande}, \bibinfo{person}{David Lo}, {and} \bibinfo{person}{Lingxiao
  Jiang}.} \bibinfo{year}{2013}\natexlab{}.
\newblock \showarticletitle{Network structure of social coding in github}. In
  \bibinfo{booktitle}{{\em Software maintenance and reengineering (csmr), 2013
  17th european conference on}}. IEEE, \bibinfo{pages}{323--326}.
\newblock


\bibitem[\protect\citeauthoryear{Tsay, Dabbish, and Herbsleb}{Tsay
  et~al\mbox{.}}{2014a}]%
        {tsay2014influence}
\bibfield{author}{\bibinfo{person}{Jason Tsay}, \bibinfo{person}{Laura
  Dabbish}, {and} \bibinfo{person}{James Herbsleb}.}
  \bibinfo{year}{2014}\natexlab{a}.
\newblock \showarticletitle{Influence of social and technical factors for
  evaluating contribution in GitHub}. In \bibinfo{booktitle}{{\em Proceedings
  of the 36th international conference on Software engineering}}. ACM,
  \bibinfo{pages}{356--366}.
\newblock


\bibitem[\protect\citeauthoryear{Tsay, Dabbish, and Herbsleb}{Tsay
  et~al\mbox{.}}{2014b}]%
        {tsay2014let}
\bibfield{author}{\bibinfo{person}{Jason Tsay}, \bibinfo{person}{Laura
  Dabbish}, {and} \bibinfo{person}{James Herbsleb}.}
  \bibinfo{year}{2014}\natexlab{b}.
\newblock \showarticletitle{Let's talk about it: evaluating contributions
  through discussion in GitHub}. In \bibinfo{booktitle}{{\em Proceedings of the
  22nd ACM SIGSOFT International Symposium on Foundations of Software
  Engineering}}. ACM, \bibinfo{pages}{144--154}.
\newblock


\bibitem[\protect\citeauthoryear{Tsay, Dabbish, and Herbsleb}{Tsay
  et~al\mbox{.}}{2012}]%
        {tsay2012social}
\bibfield{author}{\bibinfo{person}{Jason~T Tsay}, \bibinfo{person}{Laura
  Dabbish}, {and} \bibinfo{person}{James Herbsleb}.}
  \bibinfo{year}{2012}\natexlab{}.
\newblock \showarticletitle{Social media and success in open source projects}.
  In \bibinfo{booktitle}{{\em Proceedings of the ACM 2012 conference on
  computer supported cooperative work companion}}. ACM,
  \bibinfo{pages}{223--226}.
\newblock


\bibitem[\protect\citeauthoryear{Van Der~Veen, Gousios, and Zaidman}{Van
  Der~Veen et~al\mbox{.}}{2015}]%
        {van2015automatically}
\bibfield{author}{\bibinfo{person}{Erik Van Der~Veen},
  \bibinfo{person}{Georgios Gousios}, {and} \bibinfo{person}{Andy Zaidman}.}
  \bibinfo{year}{2015}\natexlab{}.
\newblock \showarticletitle{Automatically prioritizing pull requests}. In
  \bibinfo{booktitle}{{\em Proceedings of the 12th Working Conference on Mining
  Software Repositories}}. IEEE Press, \bibinfo{pages}{357--361}.
\newblock


\bibitem[\protect\citeauthoryear{Vasilescu, Van~Schuylenburg, Wulms,
  Serebrenik, and van~den Brand}{Vasilescu et~al\mbox{.}}{2014}]%
        {vasilescu2014continuous}
\bibfield{author}{\bibinfo{person}{Bogdan Vasilescu}, \bibinfo{person}{Stef
  Van~Schuylenburg}, \bibinfo{person}{Jules Wulms}, \bibinfo{person}{Alexander
  Serebrenik}, {and} \bibinfo{person}{Mark~GJ van~den Brand}.}
  \bibinfo{year}{2014}\natexlab{}.
\newblock \showarticletitle{Continuous integration in a social-coding world:
  Empirical evidence from GitHub}. In \bibinfo{booktitle}{{\em Software
  Maintenance and Evolution (ICSME), 2014 IEEE International Conference on}}.
  IEEE, \bibinfo{pages}{401--405}.
\newblock


\bibitem[\protect\citeauthoryear{Vinayak and Hassibi}{Vinayak and
  Hassibi}{2016}]%
        {vinayak16}
\bibfield{author}{\bibinfo{person}{R.K. Vinayak} {and} \bibinfo{person}{B.
  Hassibi}.} \bibinfo{year}{2016}\natexlab{}.
\newblock \showarticletitle{Clustering by Comparison: Stochastic Block Model
  for Inference in Crowdsourcing}. In \bibinfo{booktitle}{{\em Workshop Machine
  Learning and Crowdsourcing}}.
\newblock


\bibitem[\protect\citeauthoryear{Wang, Ipeirotis, and Provost}{Wang
  et~al\mbox{.}}{2013}]%
        {wang2013quality}
\bibfield{author}{\bibinfo{person}{Jing Wang}, \bibinfo{person}{Panagiotis~G
  Ipeirotis}, {and} \bibinfo{person}{Foster Provost}.}
  \bibinfo{year}{2013}\natexlab{}.
\newblock \showarticletitle{Quality-based pricing for crowdsourced workers}.
\newblock  (\bibinfo{year}{2013}).
\newblock


\bibitem[\protect\citeauthoryear{Yin, Chen, and Sun}{Yin et~al\mbox{.}}{2014}]%
        {yin2014monetary}
\bibfield{author}{\bibinfo{person}{Ming Yin}, \bibinfo{person}{Yiling Chen},
  {and} \bibinfo{person}{Yu-An Sun}.} \bibinfo{year}{2014}\natexlab{}.
\newblock \showarticletitle{Monetary interventions in crowdsourcing task
  switching}. In \bibinfo{booktitle}{{\em Second AAAI Conference on Human
  Computation and Crowdsourcing}}.
\newblock


\bibitem[\protect\citeauthoryear{Yu, Wang, Filkov, Devanbu, and Vasilescu}{Yu
  et~al\mbox{.}}{2015}]%
        {yu2015wait}
\bibfield{author}{\bibinfo{person}{Yue Yu}, \bibinfo{person}{Huaimin Wang},
  \bibinfo{person}{Vladimir Filkov}, \bibinfo{person}{Premkumar Devanbu}, {and}
  \bibinfo{person}{Bogdan Vasilescu}.} \bibinfo{year}{2015}\natexlab{}.
\newblock \showarticletitle{Wait for it: Determinants of pull request
  evaluation latency on GitHub}. In \bibinfo{booktitle}{{\em 2015 IEEE/ACM 12th
  Working Conference on Mining Software Repositories}}. IEEE,
  \bibinfo{pages}{367--371}.
\newblock


\bibitem[\protect\citeauthoryear{Yu, Wang, Yin, and Ling}{Yu
  et~al\mbox{.}}{2014a}]%
        {yu2014reviewer}
\bibfield{author}{\bibinfo{person}{Yue Yu}, \bibinfo{person}{Huaimin Wang},
  \bibinfo{person}{Gang Yin}, {and} \bibinfo{person}{Charles~X Ling}.}
  \bibinfo{year}{2014}\natexlab{a}.
\newblock \showarticletitle{Reviewer Recommender of Pull-Requests in GitHub.}
\newblock \bibinfo{journal}{{\em ICSME\/}}  \bibinfo{volume}{14}
  (\bibinfo{year}{2014}), \bibinfo{pages}{610--613}.
\newblock


\bibitem[\protect\citeauthoryear{Yu, Wang, Yin, and Ling}{Yu
  et~al\mbox{.}}{2014b}]%
        {yu2014should}
\bibfield{author}{\bibinfo{person}{Yue Yu}, \bibinfo{person}{Huaimin Wang},
  \bibinfo{person}{Gang Yin}, {and} \bibinfo{person}{Charles~X Ling}.}
  \bibinfo{year}{2014}\natexlab{b}.
\newblock \showarticletitle{Who should review this pull-request: Reviewer
  recommendation to expedite crowd collaboration}. In \bibinfo{booktitle}{{\em
  2014 21st Asia-Pacific Software Engineering Conference}},
  Vol.~\bibinfo{volume}{1}. IEEE, \bibinfo{pages}{335--342}.
\newblock


\bibitem[\protect\citeauthoryear{Zhang, Yin, Yu, and Wang}{Zhang
  et~al\mbox{.}}{2014}]%
        {zhang2014investigating}
\bibfield{author}{\bibinfo{person}{Yang Zhang}, \bibinfo{person}{Gang Yin},
  \bibinfo{person}{Yue Yu}, {and} \bibinfo{person}{Huaimin Wang}.}
  \bibinfo{year}{2014}\natexlab{}.
\newblock \showarticletitle{Investigating social media in GitHub's
  pull-requests: a case study on Ruby on Rails}. In \bibinfo{booktitle}{{\em
  Proceedings of the 1st International Workshop on Crowd-based Software
  Development Methods and Technologies}}. ACM, \bibinfo{pages}{37--41}.
\newblock


\end{thebibliography}

\end{document}